\documentclass[12pt]{iopart}

\pdfoutput=1

\usepackage{fancyhdr}

\usepackage{bm}
\usepackage[english]{babel}
 \expandafter\let\csname equation*\endcsname\relax
 \expandafter\let\csname endequation*\endcsname\relax
\usepackage{amsmath,
amssymb,amsfonts,latexsym}
\usepackage{graphicx}
\usepackage{color}
\usepackage{iopams}
\usepackage{subfigure}
\usepackage{afterpage}
\usepackage{blkarray}
\usepackage{cite}
\usepackage{tikz}
\usetikzlibrary{snakes}

\usepackage{xcolor}
\usepackage{braket}
\usepackage{float}

\usepackage{times}
\usepackage[colorlinks,citecolor=blue,linkcolor=red,urlcolor=blue]{hyperref}

\usepackage{cancel}

\newcommand{\be}{\begin{equation}}
\newcommand{\ee}{\end{equation}}
\newcommand{\bea}{\begin{eqnarray}}
\newcommand{\eea}{\end{eqnarray}}

\usepackage{etoolbox}

\makeatletter
\def\@mkboth#1#2{}
\newlength\appendixwidth
\preto\appendix{\addtocontents{toc}{\protect\patchl@section}}
\newcommand{\patchl@section}{%
  \settowidth{\appendixwidth}{\textbf{Appendix }}%
  \addtolength{\appendixwidth}{1.5em}%
  \patchcmd{\l@section}{1.5em}{\appendixwidth}{}{\ddt}%
}

\begin{document} 

\title{Entanglement and relative entropies for low-lying excited states in inhomogeneous  one-dimensional quantum systems} 
\author{Sara Murciano$^{1}$, Paola Ruggiero$^{1}$ and Pasquale Calabrese$^{1,2}$}
\address{$^1$ SISSA and INFN, via Bonomea 265, 34136 Trieste, Italy}
\address{$^2$ International Centre for Theoretical Physics (ICTP), I-34151, Trieste, Italy}
\date{\today}

\begin{abstract}
Conformal field theories in curved backgrounds have been used to describe inhomogeneous one-dimensional systems, 
such as quantum gases in trapping potentials and non-equilibrium spin chains. 
This approach provided, in a elegant and simple fashion, non-trivial analytic predictions for quantities, such as the entanglement entropy, that are not 
accessible through other methods. 
Here,  we generalise this approach to low-lying excited states, focusing on the entanglement and relative entropies in an inhomogeneous free-fermionic system.  
Our most important finding is that the universal scaling function characterising these entanglement measurements is the same  
as the one for homogeneous systems, but expressed in terms of a different variable.
This new scaling variable is a non-trivial function of the subsystem length and system's inhomogeneity that is easily written in terms of the curved metric. 
We test our predictions against exact numerical calculations in the free Fermi gas trapped by a harmonic potential, finding perfect agreement.
\end{abstract}

\setlength\parindent{0pt}

\maketitle

\section{Introduction}

The characterisation of the entanglement content of an extended quantum system is nowadays a widely studied research theme \cite{intro1, intro2, intro3}. 
The interest in quantifying entanglement in quantum systems comes in fact from different communities, ranging from  quantum information, to 
high-energy--in particular quantum gravity--, to get to condensed-matter: it is now understood that many-body quantum systems 
and their phases may be characterised through their entanglement properties.

Arguably, for a pure state $|\Psi\rangle$, the most studied quantity in this context is the entanglement entropy between two complementary spatial regions, $A$ and $ \bar{A}$, 
of an extended quantum system \cite{intro2,intro3} 
\begin{equation}
S_A = -\text{Tr} \left( \rho_A \log \rho_A \right),
\end{equation}
where $\rho_A={\rm Tr}_{\bar A} |\Psi\rangle\langle\Psi|$ is the reduced density matrix (RDM) associated to the subsystem $A$. 
Also, R\'enyi entropies
\begin{equation}
S_{A}^{(n)}= \frac{1}{1-n} \log \text{Tr} \rho_A^{n},
\end{equation}
are widely used entanglement measures \cite{intro0}.
Given that the von Neumann entropy is obtained as the limit for $n\to 1$ of $S_{A}^{(n)}$, 
the R\'enyi entropies are a very useful theoretical tool, being the core of the replica approach to the entanglement  \cite{cc-04,cc-09}.
Furthermore, they can be measured, for integer $n\geq2$, both in Monte Carlo simulations \cite{mc} and in real experiments \cite{daley-2012,lukin-18,evdcz-18,exp1, exp2, exp3,bej-18};
finally their knowledge provides information about the entire spectrum of the RDM \cite{cl-08}. 

In gapped phases the entanglement entropies satisfy the \emph{area law} \cite{area, srednicki,bombelli,h-07}:
the ground-state entanglement is not extensive but grows as the area of the boundary surface separating the subsystem $A$ from its complement.
Conversely, in gapless systems there may be universal violations to the area law.
The most remarkable  example is represented by one-dimensional systems whose low-energy physics is captured by a Conformal Field Theory (CFT) \cite{hlw-94, vidal}: 
in this case, when $A$ is finite interval of length $\ell$ embedded in a ring of length $L$, the entanglement grows as \cite{cc-04, cc-09}
\begin{equation}
S_{A}^{(n)}= \frac{c}6 \Big(1+\frac{1}{n}\Big) \log\Big(\frac{L}{\pi}\sin \frac{\pi\ell}L\Big)  +s_n,
\end{equation}
where $c$ is the central charge of the underlying CFT and is a very important {\it universal} feature of critical systems.
The additive constant $s_n$ is instead non-universal and depends on the fine details of the considered model.

It has been pointed out that not only the ground state entanglement entropy presents universal aspects but also all low-energy eigenstates \cite{sierra,afc-09}.
The R\'enyi entanglement entropy of these excitations, when corresponding to conformal primary fields, have been characterised in Refs. \cite{sierra,berganza};
it has been shown that they are  related to conformal properties of the operator defining the targeted excitation by some universal functions. 
In fact, denoting by $\rho_{\Upsilon}$ the  RDM of the state associated to the field $\Upsilon$ (and hence $\rho_{\mathbb{I}}$ is the ground-state RDM)
the ratio 
\begin{equation} 
\label{eq:univ0}
F_{\Upsilon, n} (A) \equiv \frac{\text{Tr} \rho_{\Upsilon}^n}{ \text{Tr} \rho_{\mathbb{I}}^n} ,
\end{equation}
is universal and calculable in CFT \cite{sierra,berganza}.
It is also useful to define 
\begin{equation} 
\label{Fhat}
\hat F_{\Upsilon, n} (A) \equiv \frac1{1-n} \log F_{\Upsilon, n} (A)=
\frac1{1-n} \log\frac{\text{Tr} \rho_{\Upsilon}^n}{ \text{Tr} \rho_{\mathbb{I}}^n} =S^{(n)}_{A, \Upsilon}- S^{(n)}_{A, GS},
\end{equation}
as the excess of R\'enyi entropy of the excited state  ($S^{(n)}_{A, \Upsilon}$) with respect to the ground state value ($S^{(n)}_{A, GS}$).

Another important object capturing the universal features of low-lying excited states is the relative entanglement entropy, 
a quantity widely studied in quantum information theory and, recently, also in the high-energy community\cite{lashkari2016,lashkari2014, casini-2016, clt-16, bekenstein-bound, hol-rel-entropy, ugajin2016, ugajin2016-2, ugajin-higherdim, abch-16,balasubramanian-14}.  
Given two RDMs, $\rho_1$ and $\rho_0$, e.g. corresponding to two low-lying excited states in a CFT, the relative entropy is defined as
\begin{equation} \label{relent}
S_{A}(\rho_1 \| \rho_0) = \text{Tr} \left( \rho_1 \log \rho_1 \right) -  \text{Tr} \left( \rho_1 \log \rho_0 \right),
\end{equation}
and can be interpreted as a measure of distinguishability of the two RDMs, providing a sort of distance between them in the Hilbert space \cite{rel1, rel2}.
In a replica approach, $S_{A}(\rho_1 \| \rho_0)$ can be obtained as the limit for $n\to1$ of the logarithm of the universal ratio \cite{lashkari2016,Paola1}
\begin{equation}
\label{eq:univrel0}
G_{n, A}(\rho_1||\rho_0) \equiv \dfrac{\mathrm{Tr}(\rho_1\rho_0^{n-1})}{\mathrm{Tr}\rho_1^n}.
\end{equation}

Unfortunately, all these universal features seem to be completely lost, at least at first sight, in real experiments where different kinds of inhomogeneities are always present. 
For example, recent important advances in cold atoms have allowed to set up experiments to measure the many-body 
entanglement  \cite{daley-2012,evdcz-18,lukin-18,exp1, exp2, exp3,bej-18,gbbs-17}.
However, these ultracold quantum gases are trapped by external (usually parabolic) potentials breaking translational invariance and therefore, a fortiori,  conformal invariance.
Nevertheless, it has been pointed out in Ref. \cite{curved} that, under certain assumptions, conformal invariance can be restored at the price of working in a curved background: 
it is still possible to have an underlying CFT description but in a spacetime  which is not flat anymore.
The key assumption for this approach is the \emph{local density approximation} (LDA), i.e. that the system exhibits separation of scales: 
there must exist a mesocopic scale $\ell$ at which the system is locally homogeneous (i.e. small compared to the scale at which inhomogeneity becomes important), 
but still contains a very large number of particles (i.e. large compared to the inter-particle distance). 
Denoting by $L$ the typical macroscopic length of the system, i.e. over which the density changes, and by $\langle \rho(x) \rangle^{-1}$ the inverse of the local average density, 
providing the microscopic scale, $\ell$ must be such that $L \ll \ell \ll \langle \rho \rangle^{-1}$.

\begin{figure}[t]
\centering
\includegraphics[height=6cm]{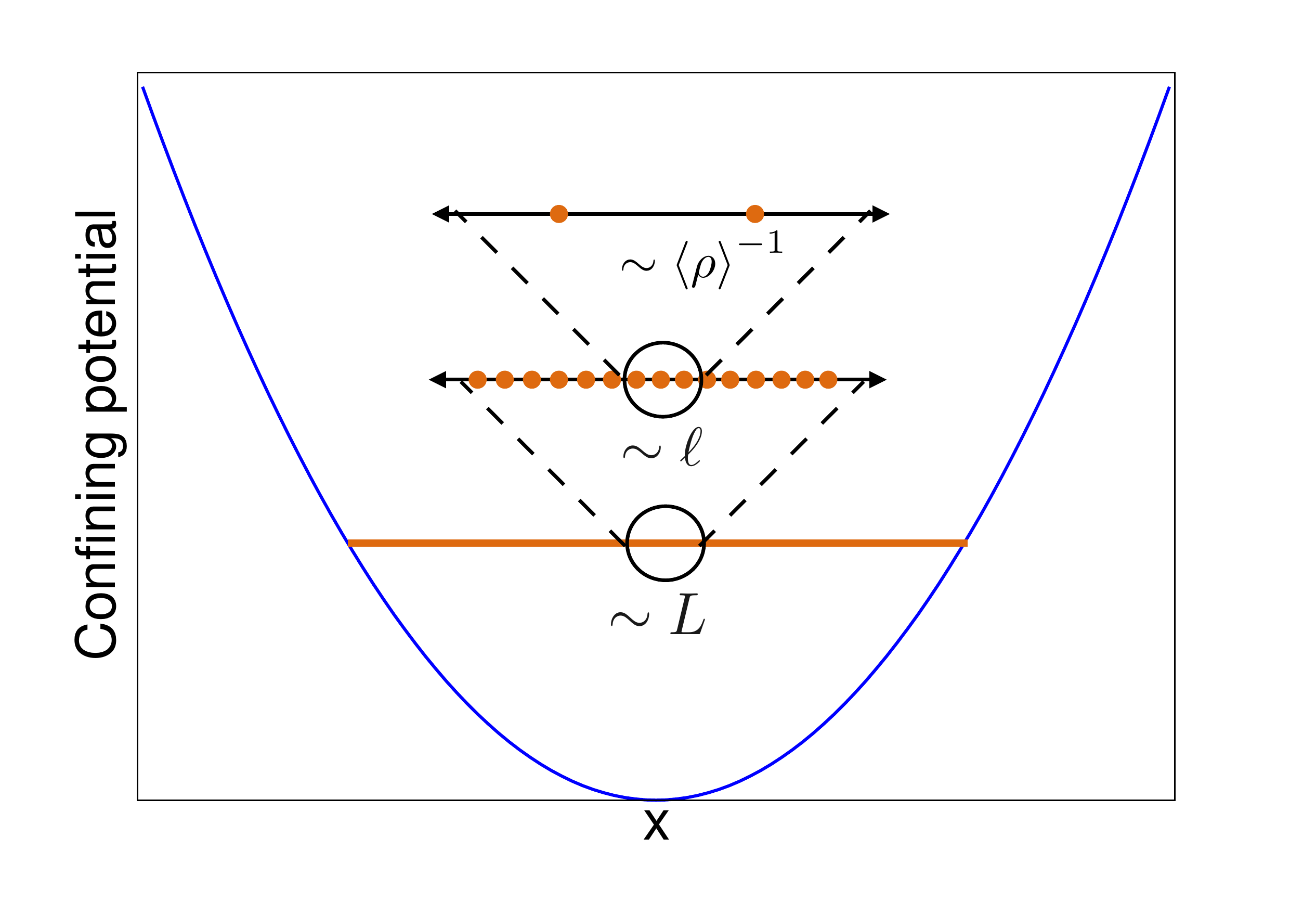}
\caption{Illustration of the local density approximation (LDA). $L$ is the typical macroscopic length of the system over which the density changes. 
The average local distance between particles is of the order of $ \langle \rho(x) \rangle ^{-1}$ ($\rho(x)$ being the density). 
The mesoscopic scale $\ell$, where the approximation holds, must satisfy $L \ll \ell \ll \langle \rho \rangle^{-1}$. }
\end{figure}

This curved CFT approach has already been employed for many applications: 
the entanglement entropies \cite{curved}, the entanglement hamiltonian  \cite{tonni-sierra},
and some correlation functions \cite{curved,bd-17} have been calculated for many different situations in inhomogeneous free-fermion models; 
the field theory description of the  rainbow model was also unveiled \cite{more-rainbow},
spin chains with gradients were studied  \cite{eisler}, and the presence of curved lightcones has  been investigated \cite{emergence}. 
All these applications refer to free models, but some results for interacting systems are also available \cite{artic,brune,gbdj-18,lm-18}.

The goal of this work is to understand whether and in which form  the universality features of the entanglement in low-lying excited states 
persist in inhomogeneous settings, as a consequence of the restored conformal invariance in curved spacetime. 
In particular we are going to focus on the ratios defined by Eqs. \eqref{eq:univ0} and  \eqref{eq:univrel0} and their analytic continuations. 
As an explicit example, we are going to employ the curved metric approach for a free Fermi gas in a (harmonic) trapping potential.
We will  show that Eqs. \eqref{eq:univ0} and  \eqref{eq:univrel0} still display very universal features. 
Interestingly and surprisingly, they turn out to be the same functions as in the analogous homogeneous setting, but of a different variable, related to the subsystem size and 
to the curved metric.


 
The manuscript is organised as follows. In Section \ref{sec:CFTcurved} we briefly review the curved CFT approach to the inhomogenous free Fermi gas 
and its bosonisation. 
Sections \ref{sec:new1} and Section \ref{sec:new2} represent the core of the paper: after recalling how the calculations of  R\'enyi and relative entropies
work in the homogeneous case, we adapt them to the inhomogeneous situation and derive the CFT predictions for different excited states.
In Section \ref{sec:nchecks} we benchmark our analytic results against exact numerical computations for Fermi gas with a a finite number $N$ of particles, 
finding excellent agreement for large $N$. 
Details about the numerical techniques are provided in \ref{appendixTools}. We finally draw our conclusions in Section \ref{sec:conclusions}.

\section{Inhomogeneous systems, CFT in curved space, and bosonisation} \label{sec:CFTcurved}

Although our approach is very general and applies to many different inhomogeneous 1D free Fermi systems, for concreteness we will focus through the entire manuscript 
on a Fermi gas trapped by an external potential. 
According to Ref.\cite{curved}, the long distance behaviour of this model is described by a massless Dirac fermion in a curved spacetime whose 
metric encodes the inhomogeneity parameters. 
We recall below the main steps to show this equivalence.

We start by considering the homogeneous free Fermi gas in 1D with hamiltonian
\begin{equation}
\label{eq:fermi}
H=\int_{-\infty}^{+\infty} dx \, c^{\dagger} (x) \left[ -\dfrac{\hbar^2}{2m} \partial^2_x -\mu \right] c(x),
\end{equation}	
being $\mu$ the chemical potential. Hereafter we set $\hbar=m=1$.
The ground-state propagator in imaginary time ($\tau=it$) is 
\begin{equation}
\label{eq:gs}
\braket{c^{\dagger}(x,\tau)c(0,0)}=\int_{-k_F}^{k_F}\dfrac{dk}{2\pi} e^{-i[kx+\varepsilon(k){\tau}]},
\end{equation} 
where $\varepsilon(k)={k^2}/{2}-\mu$ is the energy and $k_F=\sqrt{2\mu}$ is the Fermi momentum. Linearising the spectrum around the Fermi points $\pm k_F$,   $\varepsilon(k)=\pm \dfrac{d\varepsilon}{dk} \Big |_{k=k_F}(k\mp k_F)=\pm v_F(k \mp k_F)$, in the limit $x, v_F\tau \gg 1/k_F$, it becomes
\begin{equation}
\label{eq:gs1}
\braket{c^{\dagger}(x,\tau)c(0,0)} 
\simeq \dfrac{i}{2\pi} \left[ \dfrac{e^{-ik_Fx}}{x+iv_F\tau}- \dfrac{e^{ik_Fx}}{x-iv_F\tau} \right].
\end{equation}
A crucial observation is that the two terms in \eqref{eq:gs1} coincide with the R/L-components ($\psi^{\dagger}_{R,L} (x,\tau)$) of a massless Dirac fermion in 
2D euclidean spacetime, noting that
\begin{equation}
\label{eq:sol}
\braket{\psi^{\dagger}_{R,L} (x,\tau)\psi_{R,L}(0,0)}=\dfrac{1}{2}\dfrac{1}{x \pm iv_F\tau}, \qquad \braket{\psi^{\dagger}_{R,L} (x,\tau)\psi_{L,R}(0,0)}=0.
\end{equation}

We now add to the hamiltonian in Eq. \eqref{eq:fermi} an external potential $V(x)$, that at first we consider harmonic $V(x)=m\omega^2x^2/2$
(again we set $m=\omega=1$). 
In the thermodynamic limit, the density profile of fermions follows the  Wigner semicircle law \cite{vivo}
 \begin{equation}
 \label{eq:wigner}
 \rho(x)=\dfrac{1}{\pi} \sqrt{2\mu -x^2},
 \end{equation}
which is different from zero only in the interval $[-L,L]$ where $L=\sqrt{2\mu}$. 
This length is related to the total number of particles $N=\int \, dx \, \rho(x)=\mu$, hence $L=\sqrt{2N}$. 
For the LDA to hold, it should exist an intermediate scale $\ell$ which is large compared to the microscopic scale but small compared to the scale on which physical quantities vary macroscopically. For large $N$ and away from the edges, one has $N^{-1/2} \sim \langle \rho \rangle^{-1} \ll \ell \ll L \sim N^{1/2}$, and hence LDA applies. 
At this scale the system can be seen locally as homogeneous, with a Fermi momentum $k_F(x)=\pi \rho(x)$ and propagator of the same form as in \eqref{eq:gs1}
 \begin{equation}
 \label{eq:gs2}
 \braket{c^{\dagger}(x+\delta x,\tau+\delta \tau)c(x,\tau)}\simeq \dfrac{i}{2\pi} \left[ \dfrac{e^{-i (k_F(x)\delta x+iv_F(x) \delta \tau)}}{\delta x +iv_F(x) \delta \tau}-  \dfrac{e^{i (k_F(x)\delta x-iv_F(x) \delta \tau)}}{\delta x -iv_F(x) \delta \tau} \right] ,
\end{equation}
where $v_F(x) = \varepsilon'(k_F(x))$.
The only consistent Dirac theory defined on the entire domain $(x,\tau) \in [-L,L] \times \mathbb{R}$ and locally having the propagator \eqref{eq:gs2}
is  a massless Dirac action in which the metric (being the only available free parameter) varies with position.
Thus, we end up in a field theory in curved spacetime, whose action in isothermal coordinates is 
\begin{equation}
\label{eq:diracaction}
S=\dfrac{1}{2\pi}\int dz\, d\overline{z} \sqrt{g} [\psi_R^{\dagger} \overleftrightarrow{ \partial_{\overline{z}}} \psi_R+\psi_L^{\dagger} \overleftrightarrow{ \partial_{z}} \psi_L],
\end{equation}
where $g=-\det(g_{\mu \nu})$, $g_{\mu \nu}$ is the metric tensor and the line element is $d^2s=e^{2\sigma}\, dz\, d \overline{z}$, which is conformally flat. 
Indeed, the propagator associated to the action \eqref{eq:gs2} is 
\begin{equation}
\label{eq:gs3}
\braket{\psi^{\dagger}_R(z+\delta z )\psi_R(z)}=\dfrac{1}{e^{\sigma}\delta z},
\end{equation}
and coincides with \eqref{eq:gs2} upon choosing
\begin{equation}
\label{eq:geomtransf}
z(x,\tau)=\arcsin \left(\dfrac{x}{L} \right) +i\tau,
\end{equation}
defined on the infinite strip $\left[ -\frac{\pi}{2}, \frac{\pi}{2}\right] \times \mathbb{R}$. 

This formalism may simply be adapted to deal with an arbitrary potential $V(x)$. In fact, even when we do not generally know the exact solution of the single-particle problem,
in the thermodynamic limit, the relevant single-particle states are the ones very high in the spectrum for which the semi-classical approximation becomes exact. 
We emphasise that, in order to obtain non-trivial results, we need to scale the potential with the number $N$ of particles, see \cite{curved} for details.
For a generic potential, the local Fermi momentum is given by
\begin{equation}
\label{eq:localkF}
k_F(x)=\sqrt{2(\mu-V(x))},
\end{equation}
and the underlying metric is given by $ds^2=dx^2+v_F(x)^2d\tau^2$. Finally, Eq. \eqref{eq:geomtransf} becomes
\begin{equation}
\label{eq:localkF1}
z(x,\tau)=\int \dfrac{dx'}{v_F(x')}+i\tau, \qquad e^{\sigma}=v_F(x),
\end{equation}
where, we recall that $v_F(x) = \varepsilon'(k_F(x))$. 
The coordinate $z$ is defined on a strip $[x_1,x_2] \times \mathbb{R}$, where $x_1$ and $x_2$ depend on $V(x)$ and $\mu$. 

%

\subsection{Bosonisation of the Dirac fermion}

By standard bosonisation techniques \cite{DiFrancesco,GNTbook}, the action of the free Dirac fermion is mapped into a free bosonic CFT with the Euclidean action
\begin{equation}
\label{eq:hamfree111}
S=\dfrac{1}{8\pi} \int \, d\tau dx [(\partial_\tau \phi)^2+(\partial_x \phi)^2],
\end{equation}
where we set the speed of the sound $v=1$.
The bosonic field is \emph{compact}, and the target space is a ring, i.e.
\begin{equation}
\label{eq:compactified}
\phi(t,x+L) \equiv \phi(t,x)+2\pi Rm.
\end{equation}
Here $m$ is the winding number of the field configuration and $R$ is the compactification radius, that in the case of the Dirac theory is  $R=1$.

In order to clarify the operator correspondence, we should also introduce the chiral vertex operator (for left and right movers)  defined as
\begin{equation}
\label{eq:vertex}
V_{\alpha}^{R,L} =: e^{ i \alpha \phi_{R,L}(x)}:,
\end{equation}
with conformal dimension given by $h_{\alpha}= \alpha^2/2$ ($\alpha \in \mathbb{R}$) and where $\phi_{L,R}(x)$ are the left and right components of the bosonic field.

By comparing the two point correlation function of vertex operators with that of left and right moving fermionic operators,  the operator correspondence between these fields is 
\begin{equation}
\label{eq:correspondance}
\psi_R(x)=\dfrac{\eta}{\sqrt{2}} V_{1}^{R}, \qquad \psi_L(x)=\dfrac{\overline{\eta}}{\sqrt{2}} V_{1}^L \,,
\end{equation} 
where the Klein factors $\eta$ and $\overline{\eta}$ ensure the anticommutation relations between $\psi_L(x)$ and $\psi_R(x)$, being anticommuting variables themselves.
This identity states that the chiral vertex operators, Eq. \eqref{eq:vertex}, correspond, in the fermionic language,  to the annihilation/creation of a fermion. 

The derivative operator $i\partial_x \phi=i\partial_x \phi_R+i\partial_x \phi_L$ can be written in fermionic language as 
\begin{equation}
\label{eq:ferm}
\psi_R^{\dagger}(x)\psi_R(x)=-\dfrac{1}{2}:i\partial_x \phi_R(x): \quad \psi_L^{\dagger}(x)\psi_L(x)=-\dfrac{1}{2}:i\partial_x \phi_L(x):.
\end{equation}

The primary fields of the theory consists just of the vertex operator $V_1$ and the derivative field $i\partial_x \phi$ \cite{DiFrancesco}. 
In fermionic language, via bosonisation they correspond to the creation of a particle and to a particle-hole excitation respectively. 

We now discuss how the vanishing of the fermionic density at the edges of the interval $[-L, L]$ (cf. Eq. \eqref{eq:wigner})
reflects on the BCs for the free compact boson (which also lies on the strip of length $2L$).
Let us consider the fermion density operator $ \rho (x)=  :c^{\dagger} (x) c (x):$ and linearise it around the two Fermi points to get
\begin{multline}
\label{eq:continuum}
\rho (x) \sim \left[  :\psi_R^{\dagger}(x)\psi_R(x):+:\psi_L^{\dagger}(x)\psi_L(x):+ 
\right.\\ \left.
:\psi_R^{\dagger}(x)\psi_L(x):e^{-2ik_Fx}+:\psi_L^{\dagger}(x)\psi_R(x):e^{2ik_Fx} \right].
\end{multline}
Using that $\psi_R(x)$ and $\psi_L(x)$ vary very slowly on the system scale, we may drop the two terms involving the rapidly oscillating factors $e^{\pm 2ik_Fx}$.
Consequently, the vanishing of the fermion density $\rho (x)$ implies
\begin{equation}
\label{eq:fermibah}
\psi_R^{\dagger}(x)\psi_R(x)=- \psi_L^{\dagger}(x)\psi_L(x).
\end{equation}
and,  from Eq.~\eqref{eq:ferm}, for the bosonic field 
\begin{equation}
\partial_x\phi(x)=\partial_x\phi_R(x)+\partial_x\phi_L(x)=0.
\label{Nbc}
\end{equation}
Then, in this boundary CFT, the operator content of theory is halved compared to the bulk (as for any CFT \cite{cardy}).


\section{Entanglement and R\'enyi entropies of excited states of inhomogeneous systems}\label{sec:new1}

In this section we apply the formalism reviewed in Section \ref{sec:CFTcurved} to a special class of excited states in the presence of spatial inhomogeneities. 
Following Refs. \cite{sierra,berganza} we focus on the excited states generated by the action of a CFT primary operator on the ground state, 
which coincide to some low-energy excitations of the free Fermi gas. 
We will use this approach to study von Neumann entropy and the R\'enyi entropies of a Fermi gas in an arbitrary external potential.



\subsection{Homogeneous systems}

The simplest class of excited states in a CFT are those generated by the action of a primary field $\Upsilon$, with scaling dimension $\Delta$, on the CFT vacuum 
(i.e. the ground state). By the operator-state correspondence, these states may be written as
\begin{equation}
|\Upsilon\rangle\equiv \Upsilon(z=-i\infty) |0\rangle.
\label{op-st}
\end{equation}
The corresponding path-integral representation of the density matrix $\rho=|\Upsilon\rangle\langle\Upsilon|$ presents two insertions of $\Upsilon$ at $\pm i\infty$.
Taking the trace over $\bar A$ to construct $\rho_A$, sews together the fields on $\bar A$ and leaves an open cut along $A$.
Hence, the moments of the RDM, ${\rm Tr}\rho_A^n$, are $2n$-point correlation functions of $\Upsilon$ on an $n$-sheeted Riemann surface $\mathcal{R}_n$ \cite{sierra}.
Simple algebra leads to a compact expression for the ratio \eqref{eq:univ0} between the moment of the RDM of a subsystem $A$ in the excited state \eqref{op-st}
 and the one in the ground state, which is  \cite{sierra,berganza}
\begin{equation}
\label{eq:univ}
F_{\Upsilon, n} (A) =\dfrac{\braket{\prod_{k=0}^{n-1} \Upsilon(w^-_k)\Upsilon^{\dagger}(w^+_k) }_{\mathcal{R}_n}}{\langle \Upsilon(w^-_0)\Upsilon_1^{\dagger}(w^+_0)\rangle^n_{\mathcal{R}_1} },
\end{equation}
where $w^{\pm}_k = \pm i \infty$ are points where the operators are inserted in the $k$-th sheet of $\mathcal{R}_n$. 
Note that the non-universal contributions (as the cutoff $\epsilon$, explicitly entering as a regulator of the theory in the numerator and the denominator) cancel, 
and hence the ratio is universal. 

Eq. \eqref{eq:univ} holds for systems with both periodic (PBC) and open boundary conditions (OBC), but the worldsheet where the theory is defined,
 i.e., the Riemann surface $\mathcal{R}_n$ is different in the two cases. 
For PBC the single sheet geometry $\mathcal{R}_1$ is the cylinder (topologically equivalent to the complex plane), whereas for OBC $\mathcal{R}_1$ is 
an infinite strip \cite{sierraboundary}. 
For the periodic case, with a sequence of conformal transformations, the $n$-sheeted surface is mapped to the complex plane where the operators $\Upsilon$ 
are inserted at the roots of unity, see \cite{berganza} for details.
Using these  conformal mappings, the ratio $F_{\Upsilon, n} (A)$ is written as \cite{berganza}
\begin{equation}
\label{eq:final}
F_{\Upsilon, n}(A)=n^{-2n\Delta}e^{i2\pi (n-1)\Delta}\dfrac{\braket{\prod_{k=0}^{n-1} \Upsilon(s^-_{k,n}) \Upsilon^{\dagger}(s^+_{k,n})}_{\mathbb{C}}}{\langle \Upsilon(s^-_{0,1}) \Upsilon^{\dagger}(s^+_{0,1}) \rangle^n_{\mathbb{C}}},
\quad
{\rm where} \quad s^{\pm}_{k, n}=e^{i \pi/{n} (\mp {\rm x} +2k) },
\end{equation}
in which we defined the  scaling  variable ${\rm x}=\ell/L$ where $\ell$ is the length of the interval $A$ and $L$ the total length of the system. 
We use the symbol ${\rm x}$ to distinguish this variable from the spatial coordinate $x$.
Thus $F_{\Upsilon, n} (A)$ turns out to be a function of ${\rm x}$ only \cite{sierra, berganza}.

The $n$-th R\'enyi entanglement entropy is then given by Eq. \eqref{Fhat}.
Since the $F_{\Upsilon, n} (A)$ is function only of ${\rm x}$ for all excitations, the additional term in the entanglement entropies is always of order $1$ in $L$, 
showing that the universal logarithmic behaviour for the ground-state  persists for all low-lying excited states in CFT.
Higher excited states, in the middle of the spectrum of microscopical models, have instead generically extensive entanglement 
entropy \cite{afc-09,wkpv-13,bam-15,ac-17,vhbr-17,vr-17,z-07,nwfs-18,dll-18,gltz-05,vhbr-18}
(see \cite{ola-18} for low-lying excitations in massive field theories). 

For the case of OBC, which is of major interest here, we consider a systems of length $2L$, i.e. the segment $[-L,L]$ and a subsystem $A$ starting from the left 
edge $A=[-L,\ell]$ (note that the length of $A$ is $L+\ell$). 
Under these circumstances there are  only minor differences compared to the  calculation in the periodic case \cite{sierraboundary}.
A sequence of conformal transformations maps the strip of width $2L$ into the unit disk. 
The  correlation functions resulting from the mapping of Eq. \eqref{eq:univ} are usually more complicated to be worked out \cite{sierraboundary}; 
the only special case being the one of interest for us, namely correlation functions of chiral operators in the compact boson with free boundary conditions \eqref{Nbc}. 
In this case, by use of image charges, we end up exactly in the correlation function \eqref{eq:final}. The scaling variale ${\rm x}$ is now
\begin{equation}
{\rm x}=\frac{|x_1-x_2|}{2L}=\frac12 +\frac{\ell}{2L}\,.
\label{xdef}
\end{equation}

\subsection{Inhomogeneous systems} \label{sec:RenInh}

We now apply the formalism of Section \ref{sec:CFTcurved} to compute the entanglement entropy of the bipartition $A=[-L, \ell ]$, $\bar A= [ \ell,  L]$.
The coordinates of the associated field theory are ($z,\overline{z}$), defined for a generic trapping potential by \eqref{eq:localkF1} (that for the harmonic case 
simplified to \eqref{eq:geomtransf}). 
Making use of such coordinates, the subsystem changes to $A'=[z_1,z_2]$, where $z_1=z(-L)$ and $z_2=z(\ell)$. In particular, for the harmonic case 
\begin{equation}
\label{eq:geomtransf1}
z_1= -\dfrac{\pi}{2}, \qquad z_2= \arcsin \Big(\dfrac{\ell}{L} \Big).
\end{equation}
In this coordinate system, the total length of the strip is $L'=\pi$.

As in Eq. \eqref{eq:univ}, the ratio of moments of RDM in Eq. \eqref{eq:univ0} is then
\begin{equation} \label{eq:Fcurved}
F^{c}_{\Upsilon, n}(A) =\dfrac{\braket{\prod_{k=0}^{n-1} \Upsilon(z^-_k)\Upsilon^{\dagger}(z^+_k) }_{\mathcal{R}_n, {\rm curved}}}{\langle \Upsilon(z^-_0)\Upsilon_1^{\dagger}(z^+_0)\rangle^n_{\mathcal{R}_1, {\rm curved}} },
\end{equation}
where $z^{\pm}_{k}= z (w^{\pm}_k) = \pm i \infty$, $\mathcal{R}_1 $ is the strip $ \mathcal{S} = [-\pi/2, \pi/2] \times \mathbb{R}$ and $\mathcal{R}_n$ the 
Riemann surface obtained by joining cyclically $n$ copies of $\mathcal{R}_1$. 
We used the superscript ``$c$'' (for ``curved'') in order to avoid confusion with the same quantity for a homogeneous system: 
the difference with \eqref{eq:univ} is that the correlation functions are evaluated in a worldsheet where the metric is not flat  ($d^2 s = e^{2 \sigma (z)} dz d\bar{z} $). 
Note  that considering the universal ratio \eqref{eq:univ0} presents also the advantage that we do not have to deal with the metric dependent cutoff $\epsilon= \epsilon (x)$, 
which, for non-uniform systems, depends on the position and is a non trivial function of the scales entering in the problem. 
In the case of the free Fermi gas, there is just one of such scales ($k_F (x)^{-1}$), thus fixing this dependence; in more complicated models, where different scales 
exist, finding such function is still an open problem.

It is now convenient to re-express the correlation in the numerator of Eq. \eqref{eq:Fcurved} through twist fields as
\begin{equation}
\label{eq:numerator}
\langle \prod_{k=0}^{n-1} \Upsilon(z^-_k,\overline{z}^-_k) \Upsilon^{\dagger}(z^+_k,\overline{z}^+_k) \rangle_{\mathcal{R}_n, {\rm curved}}=\dfrac{\langle  \mathcal{T}_n(z_2)\tilde{\Upsilon} \tilde{\Upsilon}^{\dagger}  \rangle_{\mathcal{R}_1,{\rm curved }}}{ \langle \mathcal{T}_n(z_2) \rangle_{\mathcal{R}_1, {\rm curved}}},
\end{equation}
where $ \tilde{\Upsilon}=\Upsilon^{\otimes n}$ corresponds to $n$ replicas of the operator $\Upsilon$ and we relied on the very definition of twist fields \cite{ccd-08,cc-09}.
A similar approach was considered in \cite{estienne} in a different context.
Thanks to this rewriting, Eq.~\eqref{eq:Fcurved} takes the form
\begin{equation}
\label{eq:rewrite}
F^{c}_{\Upsilon, n} (A)=\dfrac{\langle \mathcal{T}_n(z_2)\tilde{\Upsilon}  \tilde{\Upsilon} ^{\dagger}  \rangle_{\mathcal{R}_1, {\rm curved}}}{\langle \mathcal{T}_n(z_2) \rangle_{\mathcal{R}_1,{\rm curved}}\langle  \Upsilon \Upsilon^{\dagger} \rangle^n_{\mathcal{R}_1, {\rm curved}}}.
\end{equation}
Since the inhomogeneity is encoded in the metric tensor, the idea is now to use a Weyl transformation to trace the calculation back to the one in flat space, as done for the entanglement entropy in the ground state in Ref. \cite{curved}.

Under a Weyl transformation, which does not act on the coordinates, but on the metric tensor only
\begin{equation}
 g_{\mu \nu}(z, \overline{z}) \rightarrow  e^{-2 \sigma(z, \overline{z})} g_{\mu \nu}(z, \overline{z}),
\end{equation}
a primary field $\phi_{\Delta} (z, \overline{z})$ of conformal dimension $\Delta$ transforms as
\begin{equation}
\phi_{\Delta}(z, \overline{z}) \rightarrow e^{-\sigma(z, \overline{z}) \Delta}\phi_{\Delta}(z, \overline{z}).
\end{equation}
%
This transformation applies to all the fields in Eq. \eqref{eq:rewrite}, therefore the prefactors due to the Weyl transformation cancel in the ratio, leading to
\begin{equation} \label{eq:Fcurved2}
F^{c}_{\Upsilon, n} (A)=\dfrac{\langle \mathcal{T}_n(z_2)\tilde{\Upsilon}  \tilde{\Upsilon} ^{\dagger}  \rangle_{\mathcal{R}_1}  }{\langle \mathcal{T}_n(z_2) \rangle_{\mathcal{R}_1} \langle  \Upsilon \Upsilon^{\dagger} \rangle^n_{\mathcal{R}_1}},
\end{equation}
where now the correlators are again evaluated on a flat space.
Moreover, since, as already mentioned, also the position-dependent cutoff $\epsilon (x)$ simplifies, Eq. \eqref{eq:Fcurved2} coincide with Eq. \eqref{eq:univ} for a homogeneous system with OBC for a system of size $L'=\pi$ and subsystem $A'= [z_1, z_2]$, i.e., 
\begin{equation}
F^{c}_{\Upsilon, n} (A) = F_{\Upsilon, n} (A').
\end{equation}

The only {\it fundamental} difference between the homogeneous and the inhomogeneous system is that, while in the former case the function $F_{\Upsilon, n} (A)$ in Eq. \eqref{eq:univ} is a function of the variable ${\rm x}= 1/2+\ell/(2L)$, in the latter $F^{c}_{\Upsilon, n} (A)$ is the same function but of the different variable
\begin{equation}
\label{eq:newcut}
{\rm x}'=\dfrac{z_2-z_1}{L'}=\dfrac{ \arcsin(-1+2{\rm x})}{\pi}+\dfrac{1}{2}.
\end{equation}
Thus we can make use of the results of Ref. \cite{sierra,sierraboundary} for the function $F_{\Upsilon , n} ({\rm x}) $ for a homogenous 
strip ($\Upsilon$ being a primary of the compact boson CFT)
to write  explicit formulas for the functions $F^{c}_{\Upsilon, n}$ in the inhomogeneous case. 
For the harmonic trapping potential these reads: 
\begin{subequations}\label{eq:nuove}
\begin{align}
F^c_{V_{\alpha} , n} ({\rm x}) &= F_{V_{\alpha} , n} ({\rm x'(x)}) =1 \label{eq:nuove11},\\
F^c_{i\partial \phi , n} ({\rm x}) &=F_{i\partial \phi , n}({\rm x'(x)})=
\left[ \left( \dfrac{2\sin(\pi {\rm x}')}{n} \right)^n \dfrac{\Gamma \left( \frac{1+n+n\csc(\pi {\rm x}')}{2} \right)}{\Gamma \left( \frac{1-n+n\csc(\pi {\rm x}')}{2} \right)} \right]^2\nonumber\\ &=
\left[ \left( \dfrac{4\sqrt{\rm x(1-x)}}{n} \right)^n \dfrac{\Gamma \left( \frac{1+n}2+\frac{n}{4\sqrt{\rm x(1-x)} } \right)}{\Gamma \left( \frac{1-n}2+\frac{n}{4\sqrt{\rm x(1-x)}} \right)} \right]^2.
\label{eq:nuove12}
\end{align}
\end{subequations}

By replica limit we get the von Neumann entanglement entropy in terms of the function (cf. Eq. \eqref{Fhat})
%
\begin{subequations}\label{analyti}
\begin{align}
\hat F^{c}_{V_{\alpha},1}({\rm x})\equiv&\lim_{n \rightarrow 1} \dfrac{1}{1-n} \log F^{c}_{V_{\alpha},n}({\rm x}) = 0,\\
\hat F^{c}_{i\partial \phi,1}({\rm x})\equiv&\lim_{n \rightarrow 1} \dfrac{1}{1-n} \log F^{c}_{i\partial \phi,n}({\rm x})=
2\log |2 \sin (\pi {\rm x}')| + 2\psi \left(\dfrac{1}{2 \sin (\pi {\rm x}')} \right)+ 2 \sin (\pi {\rm x}') 
\nonumber\\ =& 2\log (4\sqrt{\rm x(1-x)})+2\psi  \left(\dfrac{1}{4\sqrt{\rm x(1-x)}} \right)+4\sqrt{\rm x(1-x)},
\end{align}
\end{subequations}
where $\psi(z)$ is the digamma function.
The analytic continuation leading to the last equation has been derived in \cite{elc-13,CEL}.
We stress that Eqs. \eqref{eq:nuove} and \eqref{analyti} as function of ${\rm x}'$ are valid for an arbitrary external potential with ${\rm x'(x)}$ obtainable from \eqref{eq:localkF1}. 
Only when using ${\rm x'(x)}$  in \eqref{eq:newcut} we specialised to the harmonic case.

\section{Relative entanglement entropies between inhomogeneous states}\label{sec:new2}

Here we generalise the results of the previous section for the R\'enyi entropies to  the computation of the relative entropy between different pairs of low-lying excitations 
of the free Fermi gas, always using the bosonised CFT.

\subsection{Homogeneous systems}

We start with a brief review of the replica approach to the relative entropy between the reduced density matrices of two excited states associated to primary fields \cite{lashkari2016}.
The relative entropy between $\rho_1$ and $\rho_0$ (associated to the primaries $\Upsilon_1$ and $\Upsilon_0$ respectively, as  in \eqref{op-st}), 
is obtained from the replica limit of the universal ratio \eqref{eq:univrel0} \cite{lashkari2016}. 
It is also useful to define the  \emph{``R\'enyi'' relative entropies} 
\begin{equation}
\label{eq:renyiratio}
S_{n, A}(\rho_1||\rho_0) \equiv \dfrac{1}{1-n} \log G_{n, A}(\rho_1||\rho_0),
\end{equation} 
from which the relative entropy is obtained through the following replica limit
\begin{equation}
S_A (\rho_1 \| \rho_0) = \lim_{n \to 1}S_{n, A}(\rho_1||\rho_0) .
\end{equation}
Actually the quantities in Eq. \eqref{eq:renyiratio} lack a few properties to be good relative entropies and alternative definitions exist, see e.g. \cite{lashkari2014,cmst-18}.
However, our aim is only to have objects that in the replica limit provide the relative entropy and hence \eqref{eq:renyiratio} are sufficient. 
The ratio in Eq. \eqref{eq:univrel0} may be expressed in terms of correlation functions as \cite{Paola1}
\begin{equation}
\label{eq:univrel}
G_{n, A}(\rho_1||\rho_0)=\dfrac{\langle \Upsilon_1(w^-_0)\Upsilon_1^{\dagger}(w^+_0) \prod_{k=1}^{n-1} \Upsilon_0(w^-_k)\Upsilon_0^{\dagger}(w^+_k) \rangle_{\mathcal{R}_{n} } \langle \Upsilon_1(w^-_0)\Upsilon_1^{\dagger}(w^+_0)\rangle^{n-1}_{\mathcal{R}_1}}{\langle   \prod_{k=1}^{n-1} \Upsilon_1(w^-_k)\Upsilon_1^{\dagger}(w^+_k)\rangle_{\mathcal{R}_n} \langle \Upsilon_0(w^-_0)\Upsilon_0^{\dagger}(w^+_0)\rangle^{n-1}_{\mathcal{R}_1}  }.
\end{equation}
For a periodic system, after mapping the Riemann surface to the complex plane and finally to the cylinder, $G_{n, A}(\rho_1 \| \rho_0)$ is 
\begin{multline}
\label{eq:univcyl}
G_{n, A}(\rho_1||\rho_0) = \\n^{2(n-1)(\Delta_1-\Delta_0)}\dfrac{\langle \Upsilon_1(t^-_{0,n})\Upsilon_1^{\dagger}(t^+_{0,n}) \prod_{k=1}^{n-1} \Upsilon_0(t^-_{k,n})\Upsilon_0^{\dagger}(t^+_{k,n}) \rangle_{\mathrm{cyl} } \langle \Upsilon_1(t^-_{0,n})\Upsilon_1^{\dagger}(t^+_{0,n})\rangle^{n-1}_{\mathrm{cyl}}}{\langle \prod_{k=0}^{n-1} \Upsilon_1(t^-_{k,n})\Upsilon_1^{\dagger}(t^+_{k,n})\rangle_{\mathrm{cyl}} \langle \Upsilon_0(t^-_{0,n})\Upsilon_0^{\dagger}(t^+_{0,n})\rangle^{n-1}_{\mathrm{cyl}}  },
\end{multline}
where $t^{\pm}_{k, n} = \pi/n (\pm {\rm x} + 2k)$ (with $k= 0, \cdots , n-1$) and $\Delta_0$ and $\Delta_1$ are the scaling dimensions of $\Upsilon_0$ and $\Upsilon_1$ respectively. 

It is straightforward to show, following exactly the same steps as for the entanglement entropies,  that also the final expression of $G_{n, A} (\rho_1 \| \rho_0)$, 
for chiral excitations in a system with OBC has exactly the same expression as in \eqref{eq:univcyl}.
Indeed the starting points on the replicated strip (i.e. Eqs. \eqref{eq:univrel} and \eqref{eq:univ}) are identical except for the specific chiral operator inserted, which anyhow 
does not matter for the mapping.

\subsection{Inhomogeneous systems}

We now apply the formalism of Section \ref{sec:CFTcurved} to compute the relative entropy of different couples of density matrices associated to the bipartition 
$A=[-L,  \ell ]$, $\bar A= [ \ell,  L]$.
By the same reasoning of Section \ref{sec:RenInh}, the ratio we wish to compute is 
\begin{equation}
\label{eq:univrelc}
G^c_{n, A}(\rho_1||\rho_0)=\dfrac{\langle \Upsilon_1(w^-_0)\Upsilon_1^{\dagger}(w^+_0) \prod_{k=1}^{n-1} \Upsilon_0(w^-_k)\Upsilon_0^{\dagger}(w^+_k) \rangle_{\mathcal{R}_{n}, {\rm curved} } \langle \Upsilon_1(w^-_0)\Upsilon_1^{\dagger}(w^+_0)\rangle^{n-1}_{\mathcal{R}_1, {\rm curved}}}{\langle   \prod_{k=1}^{n-1} \Upsilon_1(w^-_k)\Upsilon_1^{\dagger}(w^+_k)\rangle_{\mathcal{R}_n, {\rm curved}} \langle \Upsilon_0(w^-_0)\Upsilon_0^{\dagger}(w^+_0)\rangle^{n-1}_{\mathcal{R}_1, {\rm curved}}  }.
\end{equation}
where, also in this case, we use the superscript ``$c$'' just to avoid confusion with the same quantity for a homogeneous system.
The calculations proceed in complete analogy too, leading to
\begin{equation} \label{eq:Gcurved}
G^c_{n, A}(\rho_1 \| \rho_0)= G_{n, A'}(\rho_1 \| \rho_0).
\end{equation}
Once again, the only difference between homogenous and inhomogeneous systems is that the  scaling variable is 
not ${\rm x}$ but ${\rm x}'$ as defined in \eqref{eq:newcut}.

\begin{figure}
\centering
\subfigure
  {\includegraphics[width=0.45\textwidth]{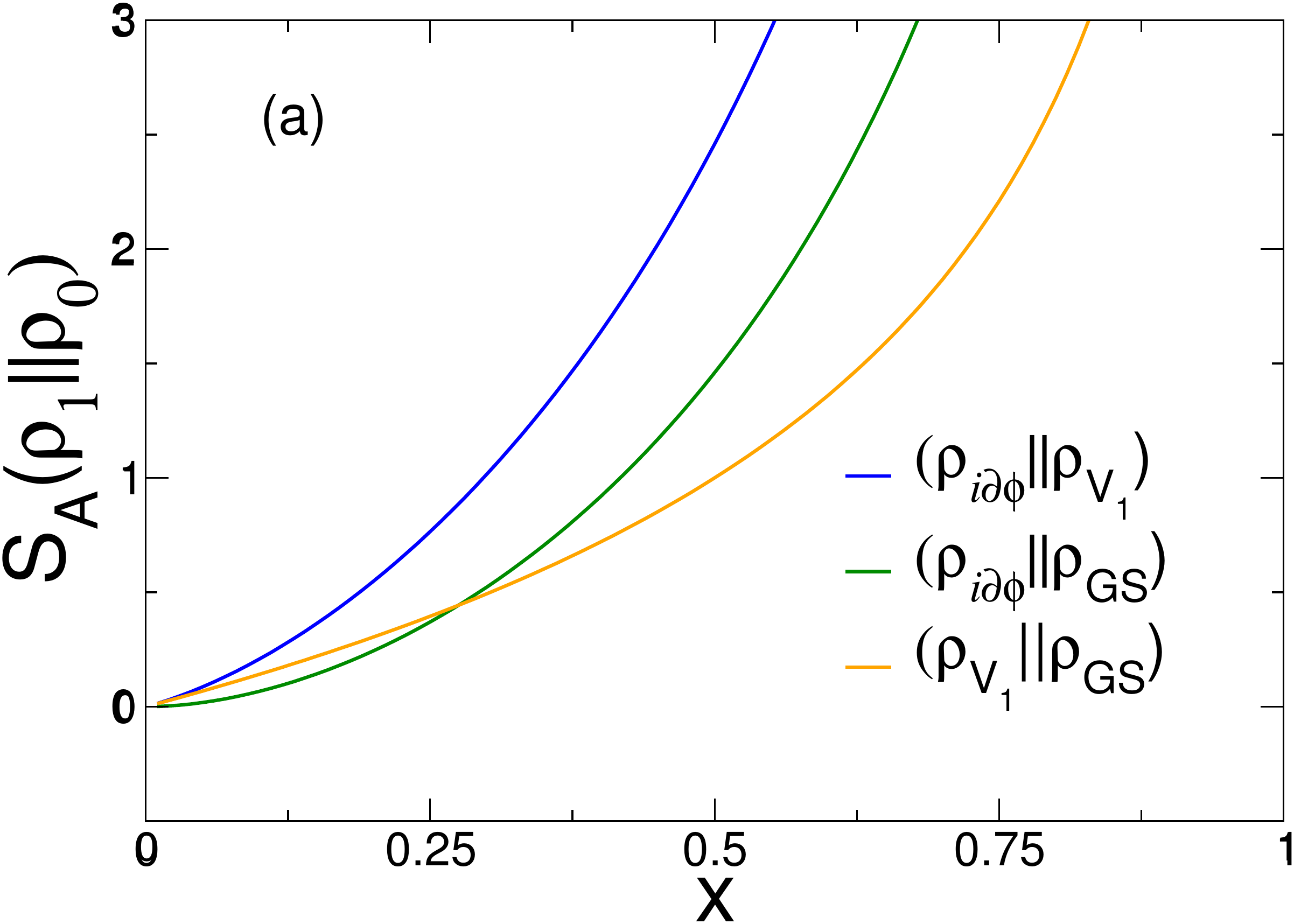}}
\subfigure
   {\includegraphics[width=0.45\textwidth]{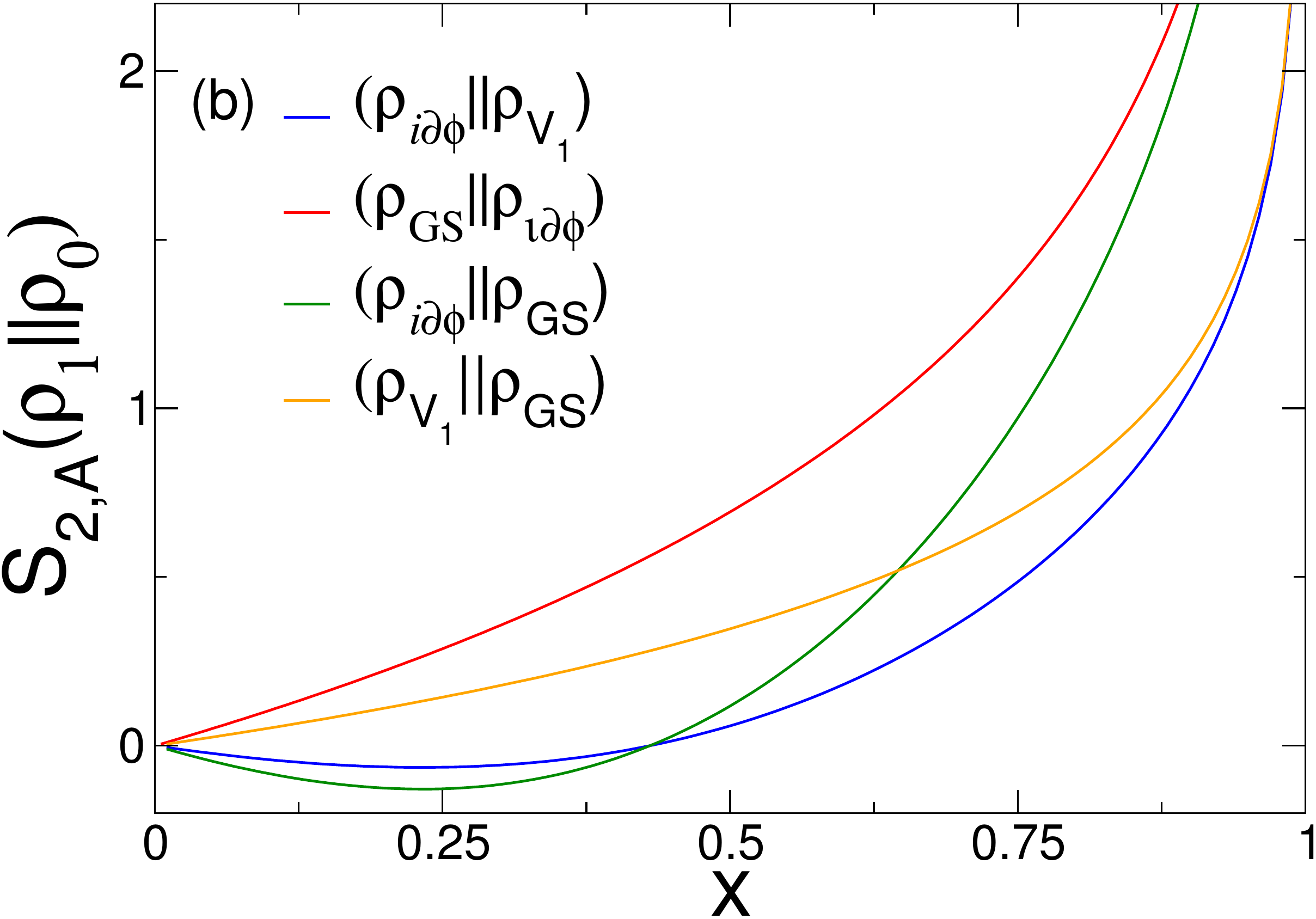}}
\subfigure
   {\includegraphics[width=0.45\textwidth]{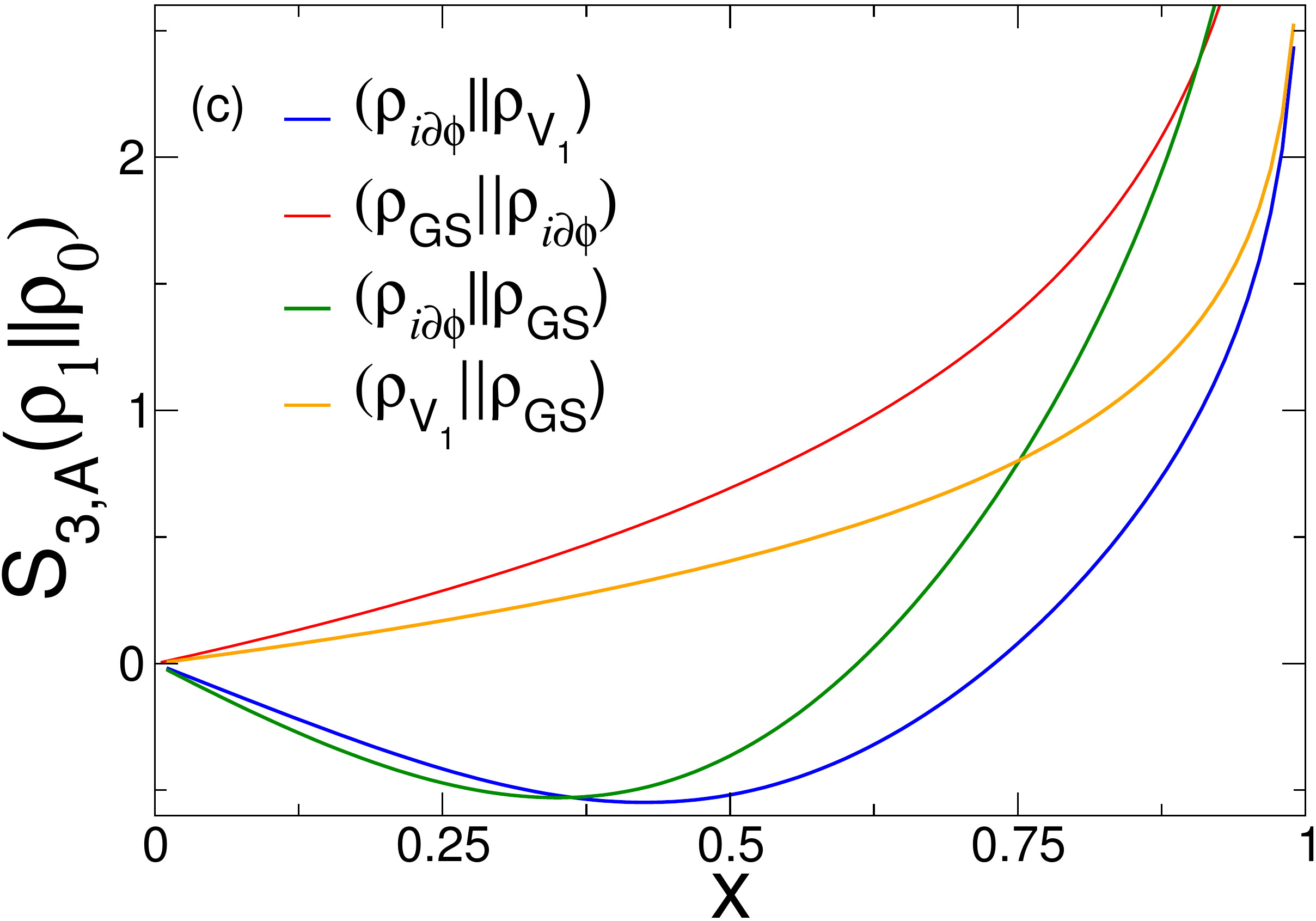}}
    \subfigure
  {\includegraphics[width=0.45\textwidth]{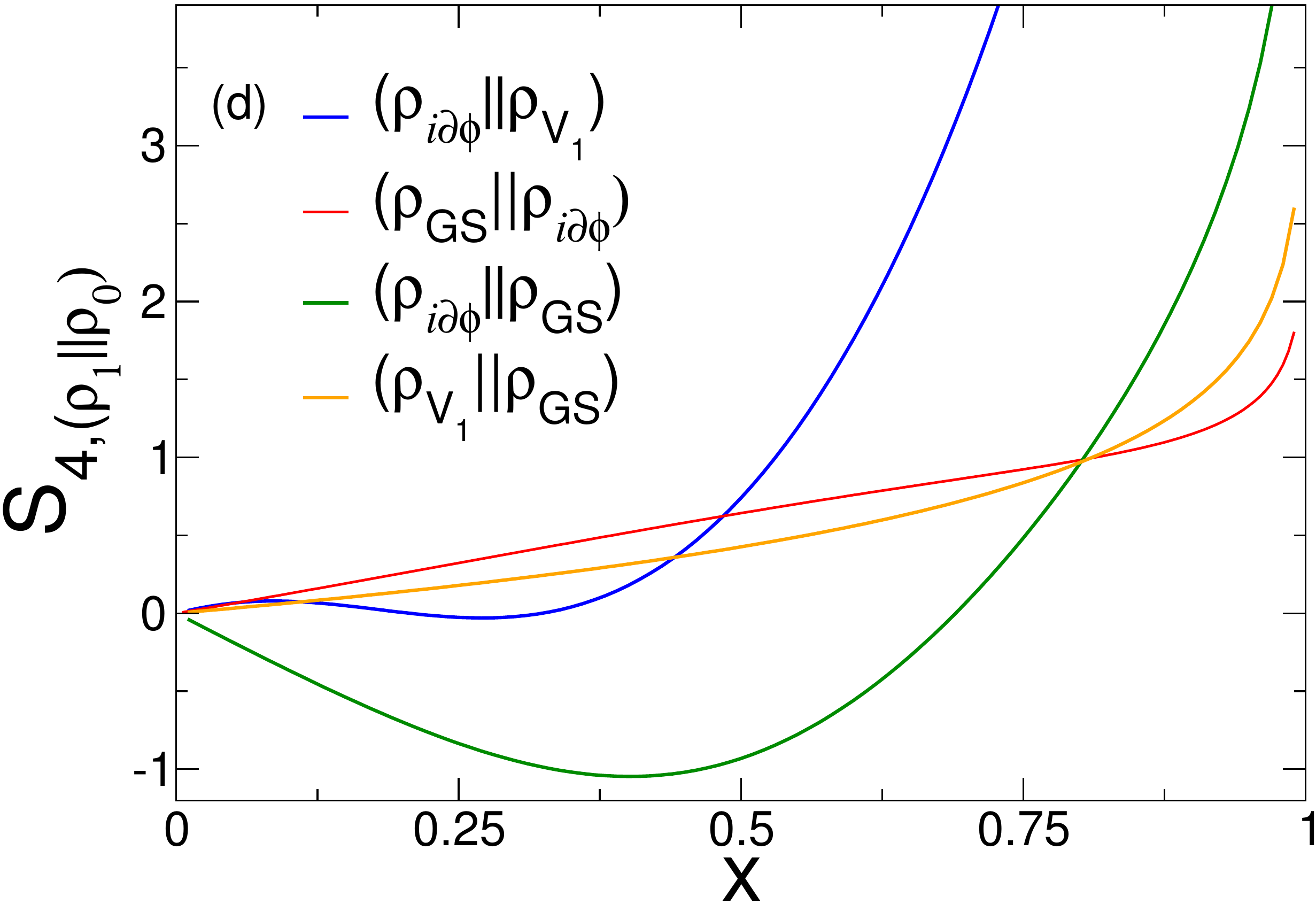}}
\caption{The CFT predictions for the R\'enyi relative entropies $S_{n, A}(\rho_1||\rho_0)$, Eq. \eqref{eq:renyiratio}, as a function of the scaling variable ${\rm x}=1/2+ \ell/(2L)$, 
for different values of $n=1,2,3, 4$ in panels (a), (b), (c) and (d) respectively. 
In each panel (at fixed $n$) we report the couples of states considered. 
}\label{fig:plotssinomogenei}
\end{figure}

Explicitly, we have  \cite{Paola1} 
\begin{subequations}\label{eq:nuoveRel}
\begin{align}
&G^c_{n,A}(\rho_{V_{\alpha}} \| \rho_{GS}) =  G^c_{n,A}(\rho_{GS}  \| \rho_{ V_{\alpha}}) =\left( \dfrac{\sin(\pi {\rm x}')}{n \sin (\pi {\rm x}' /n)} \right)^{\alpha^2},\label{eq:sseconda} \\
&G^c_{n, A}(\rho_{i\partial \phi} \| \rho_{GS})= \left( \dfrac{\sin \pi {\rm x}'}{n}\right)^{2(1-n)} 4^{-n}\sin^2 \left( \frac{\pi {\rm x}'}{n} \right)\dfrac{\Gamma^2\left( \frac{1-n+n\csc(\pi {\rm x}')}{2} \right)}{\Gamma^2\left( \frac{1+n+n\csc (\pi {\rm x}')}{2} \right)},\label{eq:pprima} \\
&G^c_{n, A}(\rho_{GS} \| \rho_{i\partial \phi} )=\left( \dfrac{\sin \pi {\rm x}'}{n}\right)^{2(n-1)} \det \left[\dfrac{1}{\sin(t_{ij} ({\rm x}')/2)} \right]_{i,j \in [1,2(n-1)]}, \label{eq:qquarta} \\
&G^c_{n, A}(\rho_{i \partial \phi} \| \rho_{V_{\beta}})= \tilde{C}_{\beta}(n,{\rm x}') G^c_{n, A}(\rho_{i\partial \phi} \| \rho_{GS}) G^c_{n,A}(\rho_{GS}  \| \rho_{ V_{\beta}})
,\label{eq:tterza}
\end{align}
\end{subequations}
where
\begin{equation}
 \tilde{C}_{\beta} (n, {\rm x}') =1 - \beta^2\sin^2 \left(\frac{\pi {\rm x}'}{n} \right) \left( \sum_{k=1}^{n-1} \cot \frac{\pi}{n}({\rm x}' + k) \right)  \left( \sum_{l=1}^{n-1} \cot \frac{\pi}{n} (-{\rm x}' + l) \right),
\end{equation}
and, by replica trick,
\begin{subequations} \label{analityRel}
\begin{align}
S_A(\rho_{GS}||\rho_{V_{\alpha}})=&S_A(\rho_{V_{\alpha}}||\rho_{GS})=\alpha^2(1-\cot (\pi {\rm x}')), \\
S_A (\rho_{i\partial \phi}||\rho_{GS})=&2 \left( \log (2\sin(\pi {\rm x}'))+1-\pi x\cot(\pi {\rm x}')+\psi\left( {\csc(\pi {\rm x}')}/{2} \right) + \sin (\pi {\rm x}')\right), \\
S_A (\rho_{i \partial \phi}|| \rho_{V_{\beta}})=&(2+\beta^2)[1-\pi {\rm x}' \cot (\pi {\rm x}')]+2\log(2(\sin (\pi {\rm x}')))+\nonumber \\ &+
2\psi \left({\csc(\pi {\rm x}')}/{2} \right)+2\sin (\pi {\rm x}').
\end{align}
\end{subequations}
All these expressions may be easily rewritten in terms of ${\rm x}$, but the final results are not very illuminating and we omit them.
As stressed in \cite{Paola1} for the calculations in the homogeneous case, it is not yet possible to derive the analytic continuation of \eqref{eq:qquarta}.

In Figure~\ref{fig:plotssinomogenei} we plot the R\'enyi relative entropy $S_{n,A}(\rho_1||\rho_0)$ for different pairs of states. 
Note that, as already observed \cite{Paola1}, while the relative entropy $S_{A}(\rho_1||\rho_0)$ is always positive and monotonous, this is not generally the case for $n\neq 1$. 
Moreover, we notice that $S_{n, A}(\rho_1 \| \rho_0)$ as a function of ${\rm x}$ goes to zero faster than in the homogeneous case.

\section{Numerical checks} \label{sec:nchecks}

We now compare the analytic formulas derived in the previous sections with the exact numerical data for the low-lying excited states of a Fermi gas trapped by a 
harmonic potential. 
The R\'enyi entropies in these excited states have been obtained by two different methods described in details in the appendices, i.e. by the
 \emph{overlap matrix} technique (\ref{appendixOM}) and by \emph{Gauss--Legendre discretisation} of the continuous correlation matrix  (\ref{appendixD}). 
Of course, they furnish identical results, providing a further test of the correctness of their implementation. 
For the calculation of the R\'enyi relative entropy the overlap matrix approach is not easily usable and therefore we only employ the Gauss--Legendre discretisation 
method for the correlation matrices and compute the trace of the product of the associated RDMs as explained in \ref{appendixGamma}.

We consider only two types of excited states. 
One is a particle-hole excitation in which the particle in the highest occupied single-particle level (i.e. the $N$-th one) is shifted up of one level (i.e. to the $(N+1)$-th one).
In the CFT, this excitation corresponds to the state generated by the action of the primary operator $i\partial \phi$.
The other corresponds to add a particle to the Fermi sea: this is a very trivial operation because the excited state is just the ground state in the canonical ensemble 
with one particle more. However, in the grand-canonical ensemble it is an excited state that in curved CFT corresponds to the action of the vertex operator 
with $\alpha=1$, i.e. $V_{1}$ (which indeed, as already mentioned, acts as creation operator of a fermion).

\begin{figure}
\centering
\subfigure
  {\includegraphics[width=0.33\textwidth]{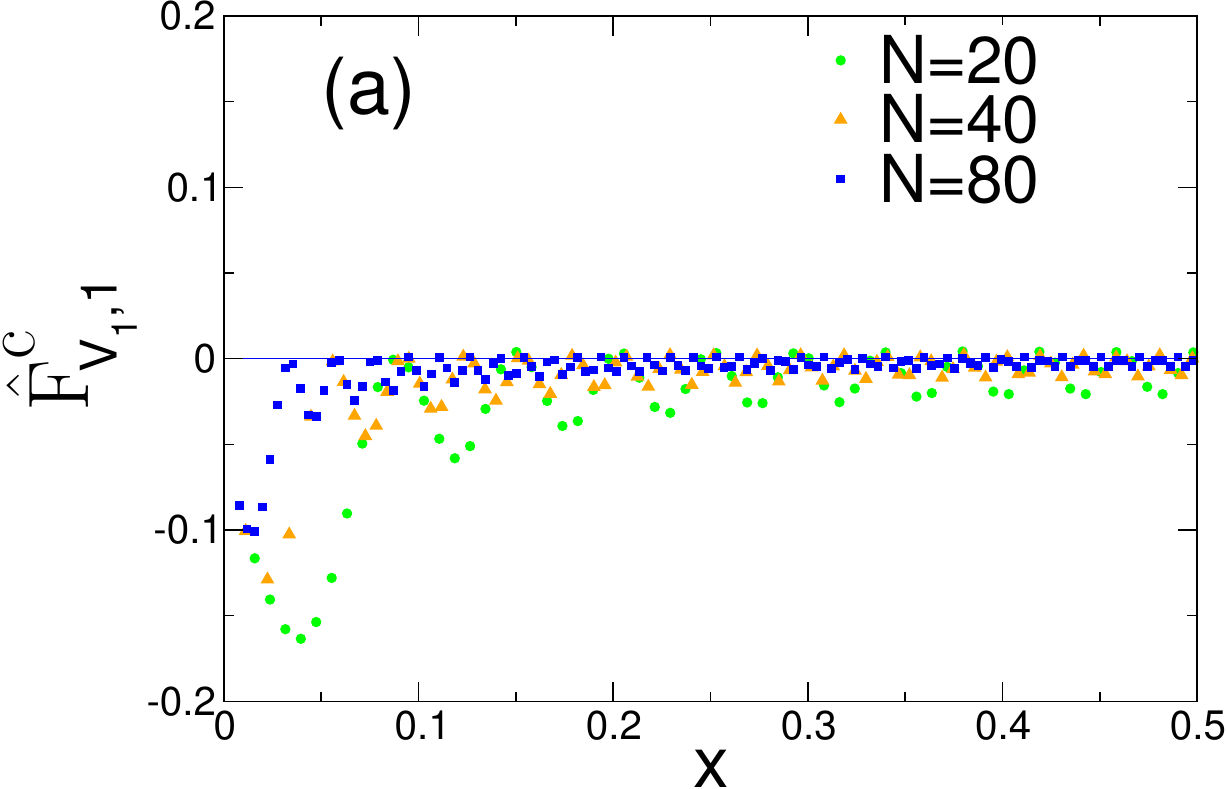}}
\subfigure
  {\includegraphics[width=0.32\textwidth]{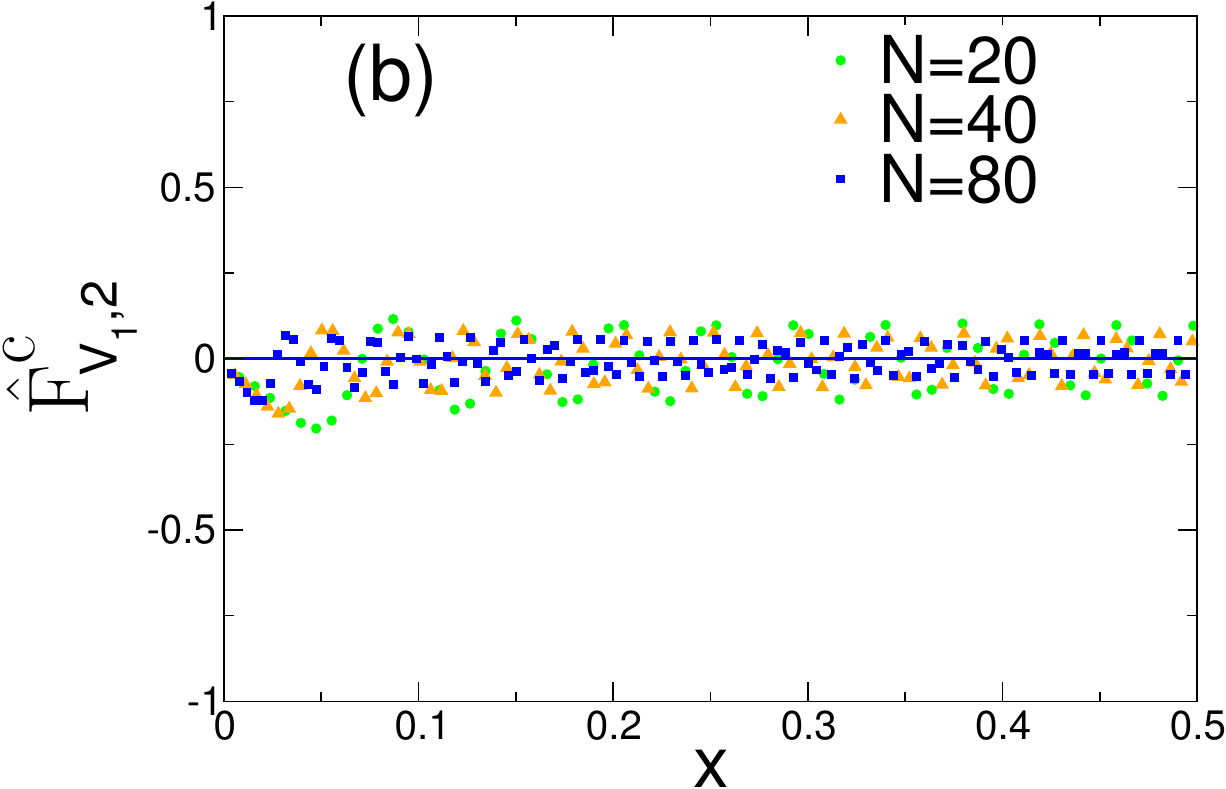}}
\subfigure
   {\includegraphics[width=0.32\textwidth]{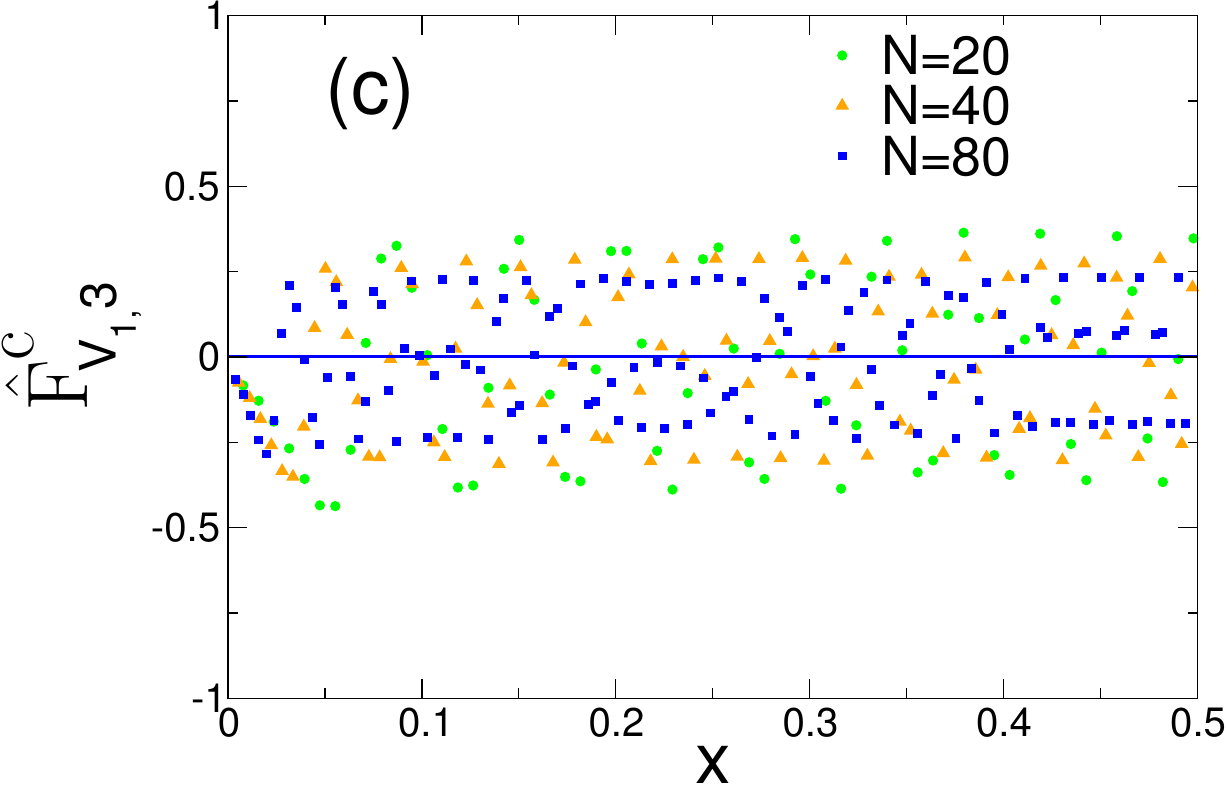}}
\caption{The universal scaling function for the R\'enyi  entanglement entropy $\hat F^c_{V_1, n}({\rm x})$ (cf. Eq.~\eqref{Fhat}), 
for the excited state given by an addition of a particle, which in CFT correspond to the vertex operator $V_1$.
Data are shown as function of ${\rm x}$ (cf. \eqref{xdef}) for different values of $n$ ($=1,2,3$ in panels (a), (b)  and (c) respectively). 
Different colours correspond to different number of particles $N$. 
The continuous curve is the CFT prediction ($\hat F^{c}_{V_1, n}({\rm x})=1$), Eq. \eqref{eq:nuove11}.
The data have been obtained with the method of the overlap matrix, \ref{appendixOM}. 
}
\label{fig:overlapVertex}
\end{figure}

\subsection{Entanglement entropies in excited states}

\begin{figure}
\centering
\subfigure
  {\includegraphics[width=0.46\textwidth]{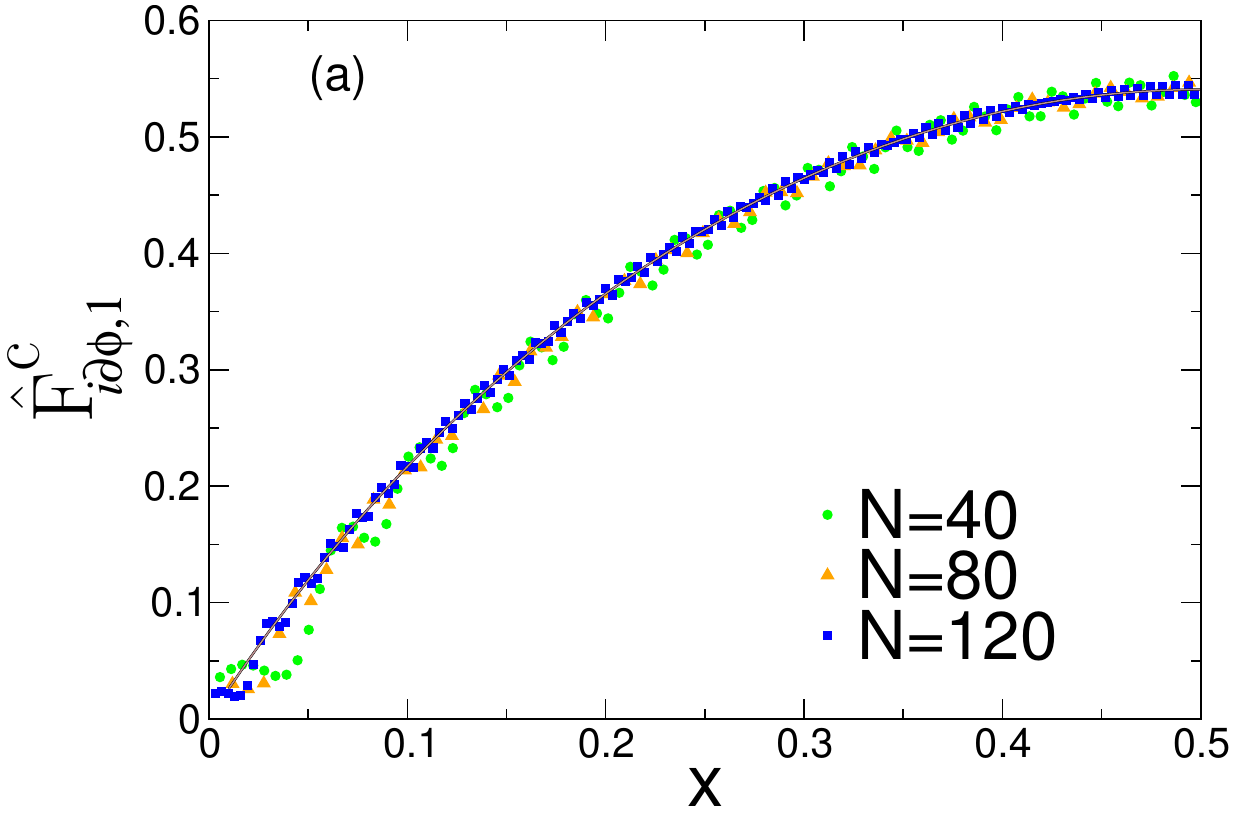}}
\subfigure
   {\includegraphics[width=0.45\textwidth]{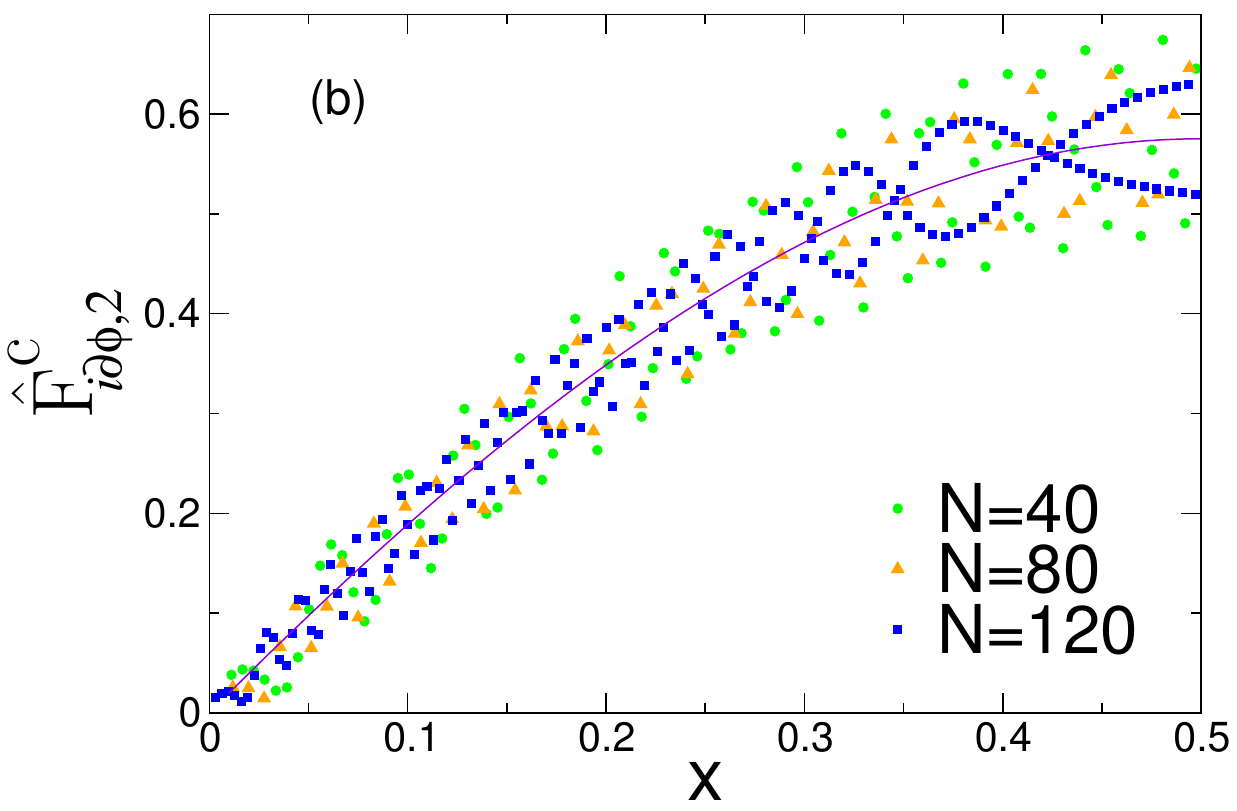}}
\subfigure
   {\includegraphics[width=0.45\textwidth]{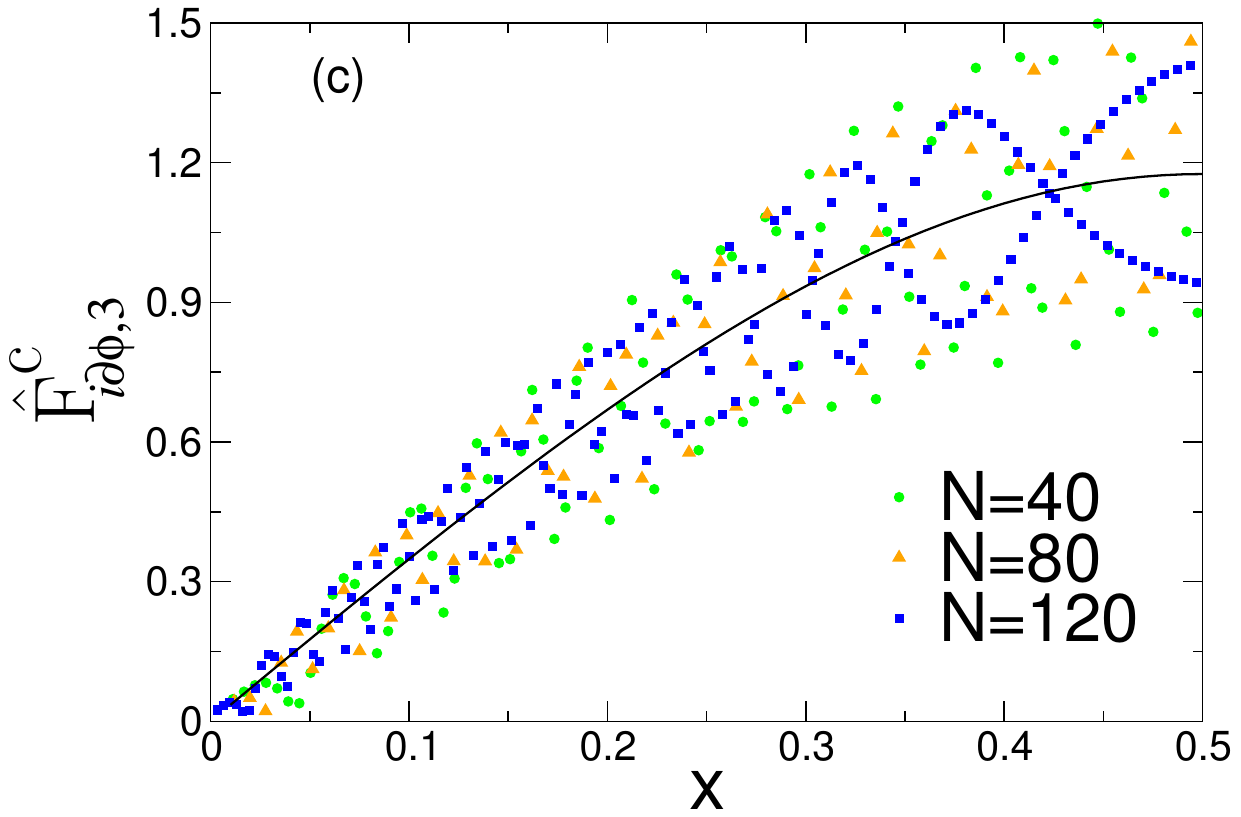}}
    \subfigure
  {\includegraphics[width=0.45\textwidth]{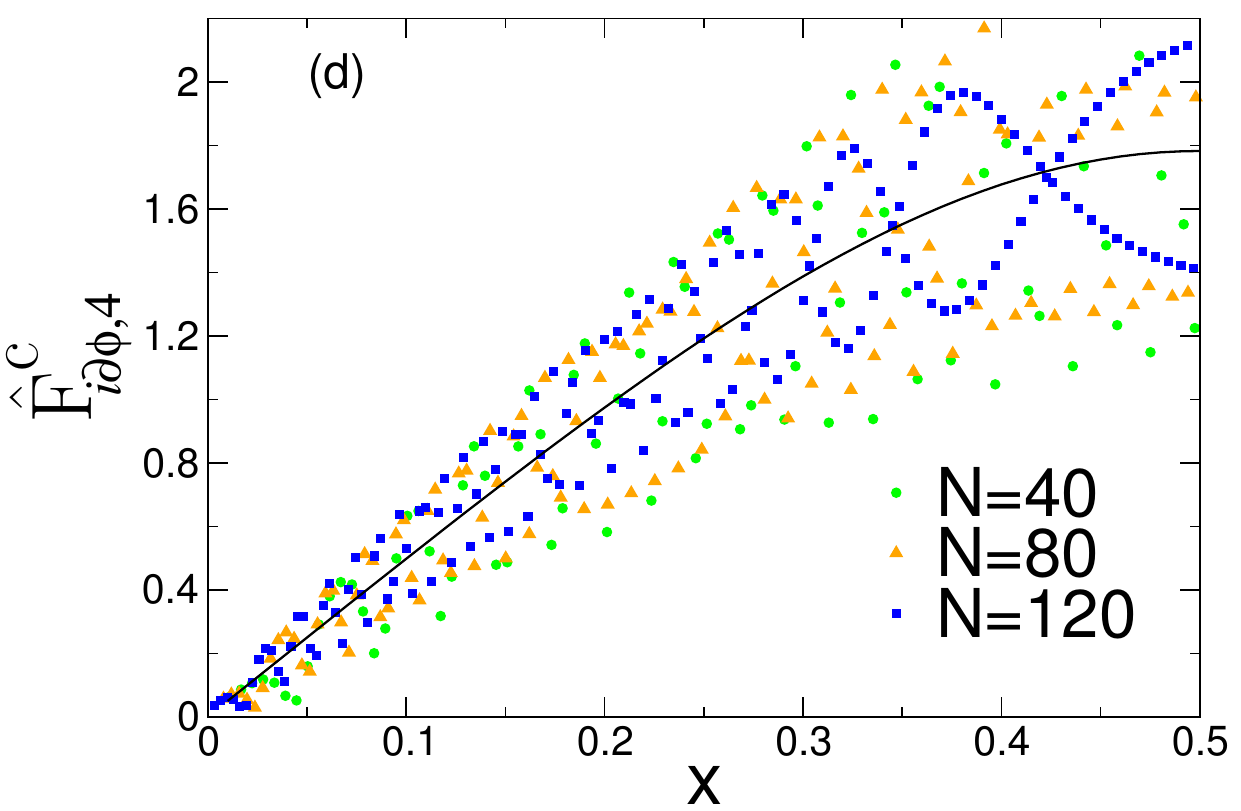}}
\caption{The universal scaling function for the R\'enyi  entanglement entropy $\hat F^c_{i\partial \phi, n}({\rm x})$, Eq. \eqref{Fhat},  for the particle-hole excitation, 
corresponding in CFT to the action of the derivative operator.
Data are shown as function of ${\rm x}$ (cf. Eq. \eqref{xdef}) for different values on $n$ ($=1,2,3, 4$ in panels (a), (b), (c) and (d) respectively). 
Different colours of the points correspond to different particle numbers $N$. The continuous curve is the CFT prediction, Eq. \eqref{eq:nuove12}.
The data have been obtained with the Gauss--Legendre discretisation method, \ref{appendixD}.
}
\label{fig:continuumm1}
\end{figure}

In Figures~\ref{fig:overlapVertex} and \ref{fig:continuumm1} we report the exact numerical data for the unversal scaling function $\hat F^{c}_{\Upsilon,n}({\rm x})$ for the 
R\'enyi entanglement entropies of a Fermi gas of $N$ particles trapped by a harmonic potential. 
In Fig. \ref{fig:overlapVertex} we report the data for the excited state with one particle more (corresponding to the vertex operator,  $\Upsilon=V_1$) 
and in Fig. \ref{fig:continuumm1} we report the data for a particle-hole excitation (corresponding to $\Upsilon=i \partial \phi$). 
In both cases, we consider the bipartition $A=[-\infty,\ell]$ and $\bar A=[\ell, \infty]$ (in the thermodynamic limit it is irrelevant whether $A$ starts 
from $-L$ or any point to its left, so we just fix to $-\infty$ which makes the numerical calculation easier). 
We plot the scaling function versus ${\rm x}=1/2+\ell/2L$ (cf. \eqref{xdef}) which is the natural variable one would have been using without knowing a priori the CFT solution in terms 
of ${\rm x}'$, cf. Eq. \eqref{eq:newcut}.
 
Let us now discuss these figures. For the excited state corresponding to the vertex operator (cf. Figure~\ref{fig:overlapVertex}),
the CFT prediction is $F_{V_1, n}({\rm x}')=1$ as in Eq. \eqref{eq:nuove11}. 
It is evident that the data converge to the CFT predictions increasing the system size, as they should.  
This is a rather trivial result since the excited state is the ground state with one particle more;
the analytic result for the ground-state entropies in the harmonic potential, cf. Refs. \cite{random,curved},  shows that the difference 
between the results at $N$ and at $N+1$ is of order $o(1)$ for large $N$.
However, in spite of this simplicity, we note the  presence of oscillating deviations from the CFT prediction which clearly decrease with system size and 
hence are subleading corrections to the scaling. These corrections get larger for larger values of the R\'enyi index $n$. 
We are going to discuss and characterise them below. 

Data for the particle-hole excited state  are  reported in Figure \ref{fig:continuumm1}, for different values of the R\'enyi index $n$ and number of particles $N$.
We also show the highly non-trivial result for the CFT excited state generated by the action of the derivative operator, cf. Eq.~\eqref{eq:nuove12}.
The data at finite $N$ are  close to the CFT predictions for all $n$ and the difference gets smaller as $N$ increases. 
In particular for $n=1$ (von Neumann entropy), the data at finite $N$ lie very close to the asymptotic prediction. 
The differences between numerical data and asymptotic predictions have to be imputed to subleading corrections
that, in analogy with the case of the vertex operator, become larger as $n$ increases. 

\begin{figure}[t]
\centering
\subfigure
  {\includegraphics[width=0.46\textwidth]{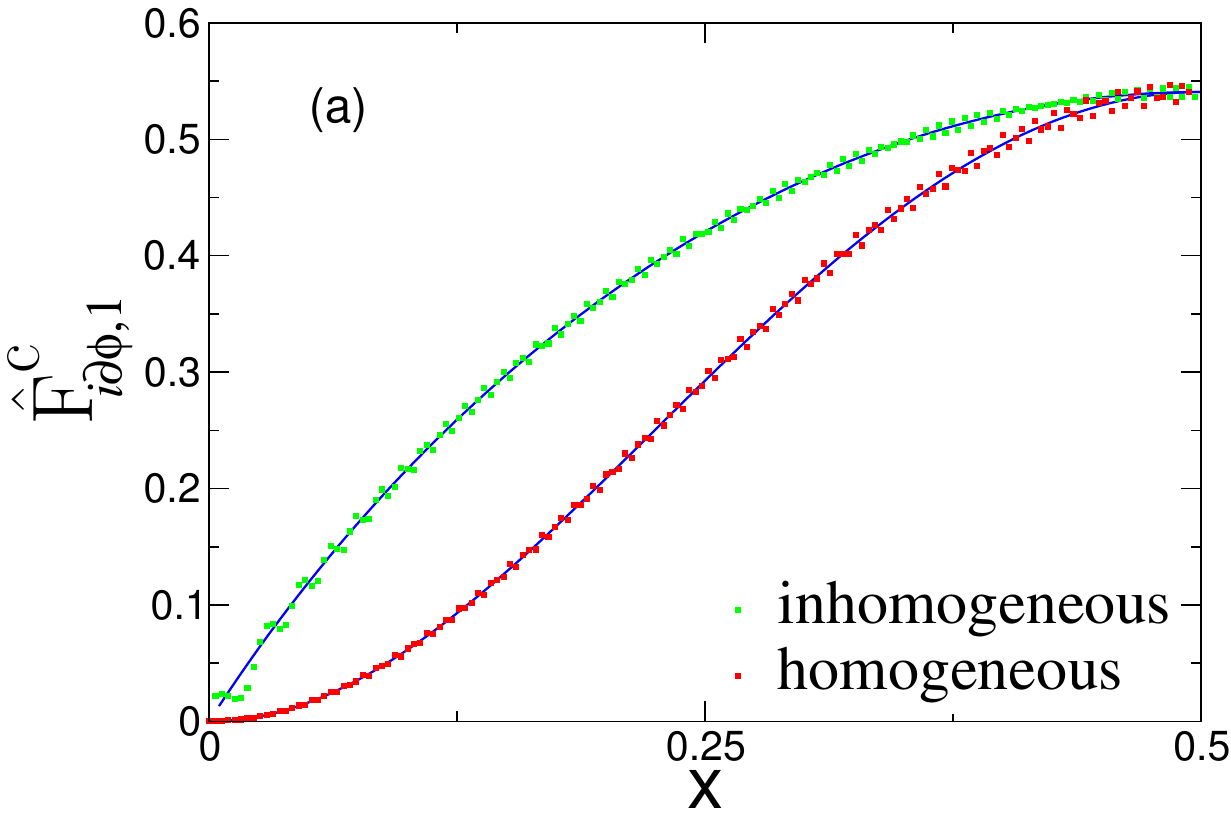}}
\subfigure
   {\includegraphics[width=0.45\textwidth]{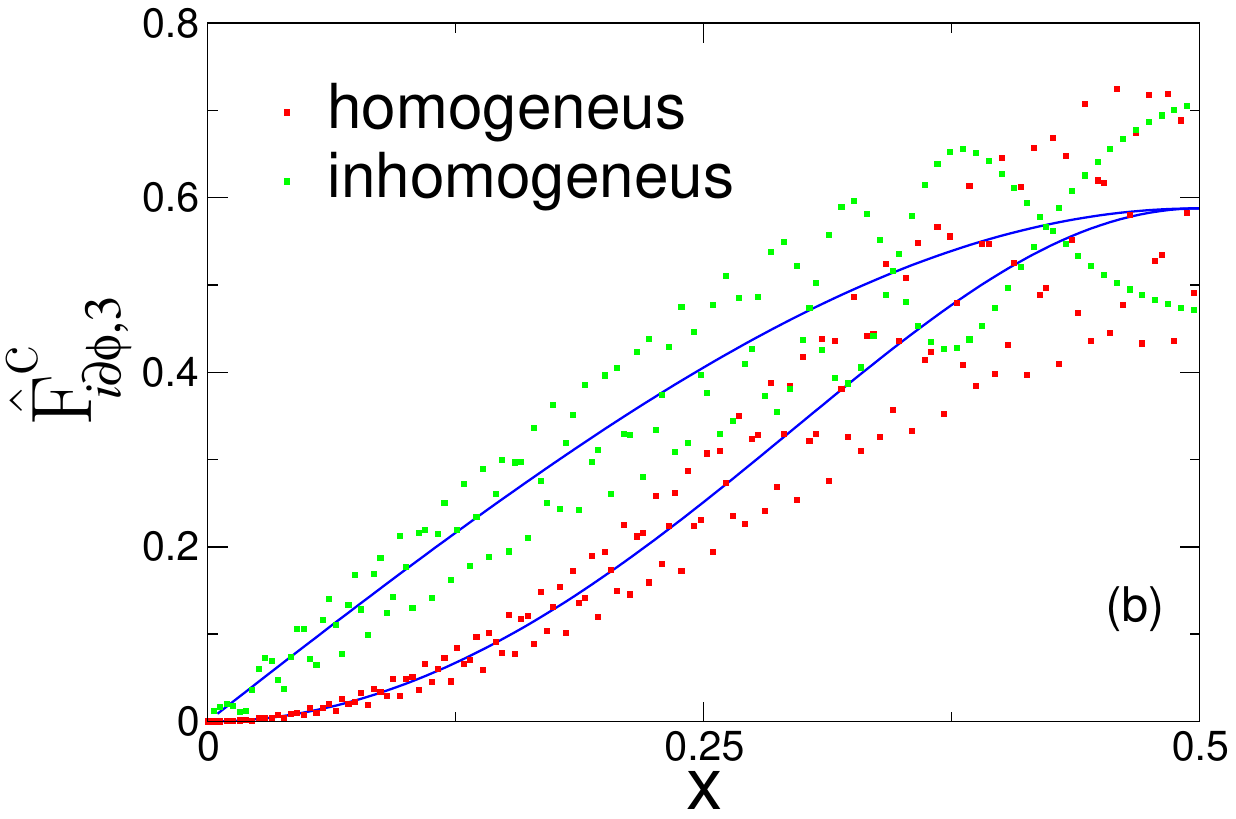}}
\caption{Comparison between the universal scaling functions $\hat F_{i\partial \phi,n}({\rm x})$ (cf. Eq. \eqref{Fhat}) for two Fermi gases: (i) with hard-wall
boundary conditions (i.e. with vanishing density at the edges $\pm L$); (ii) trapped by a harmonic potential.  
The continuous curves are the two asymptotic CFT predictions versus ${\rm x}=1/2+\ell/(2L)$ (cf. \eqref{xdef}).  
The dots correspond to exact numerical data for gases with  $N=120$ particle.
}
\label{fig:continuumm2}
\end{figure}

Before addressing these corrections to the scaling, we wish to discuss qualitative and quantitative differences between 
the scaling functions  in homogeneous and trapped systems. 
The data for $n=1$ and $n=3$ are reported once again in Fig. \ref{fig:continuumm2} together with the homogenous results for a system with 
open boundary conditions. The qualitative shapes of the two sets of curves are very different. 
In the homogenous case, the entropy excess starts off linearly at ${\rm x}=0$, while only quadratically in the trapped setting. 
Physically this behaviour may be understood as a consequence of the fact that the single-particle states at the edge have a lower density in the trapped case. 
Mathematically instead it just follows from the mapping \eqref{eq:newcut} between ${\rm x}$ and ${\rm x}'$ which is quadratic close to ${\rm x}=0$.
Furthermore, the linearity of the homogeneous scaling function for small ${\rm x}$ is a very general feature and the linear slope is proportional to the scaling dimension 
of the operator $\Upsilon$ \cite{sierra}. Hence, we conclude that for a general excited state $|\Upsilon\rangle$ in this inhomogeneous setting, 
the function $\hat F^c({\rm x})$ will start off quadratically with an amplitude proportional to the dimension of the operator $\Upsilon$.
Another interesting observation is that while the  two functions (homogenous and inhomogeneous ones) are very different, they have 
exactly the same value at ${\rm x}=1/2$, i.e. in the center of the trap.
Again, mathematically this just follows from the trivial observation that the point ${\rm x}=1/2$ is a fixed point of the map between ${\rm x}$ and ${\rm x}'$ (cf. Eq. \eqref{eq:newcut})
and hence this result is valid for an arbitrary state. 
At first, this may seem surprising for the entanglement entropy which depends on all the points to the left of the center (i.e. with $x<0$)
which are sensitive to the trapping potential. 
However, if we think to the (R\'enyi) entropy as a local one-point function of a twist field, the equivalence in the two geometries follows from the fact that 
at the center the gradient of the density vanishes in both cases and the system looks like homogenous and uniform.

\begin{figure}[t]
\centering
\subfigure
  {\includegraphics[width=0.46\textwidth]{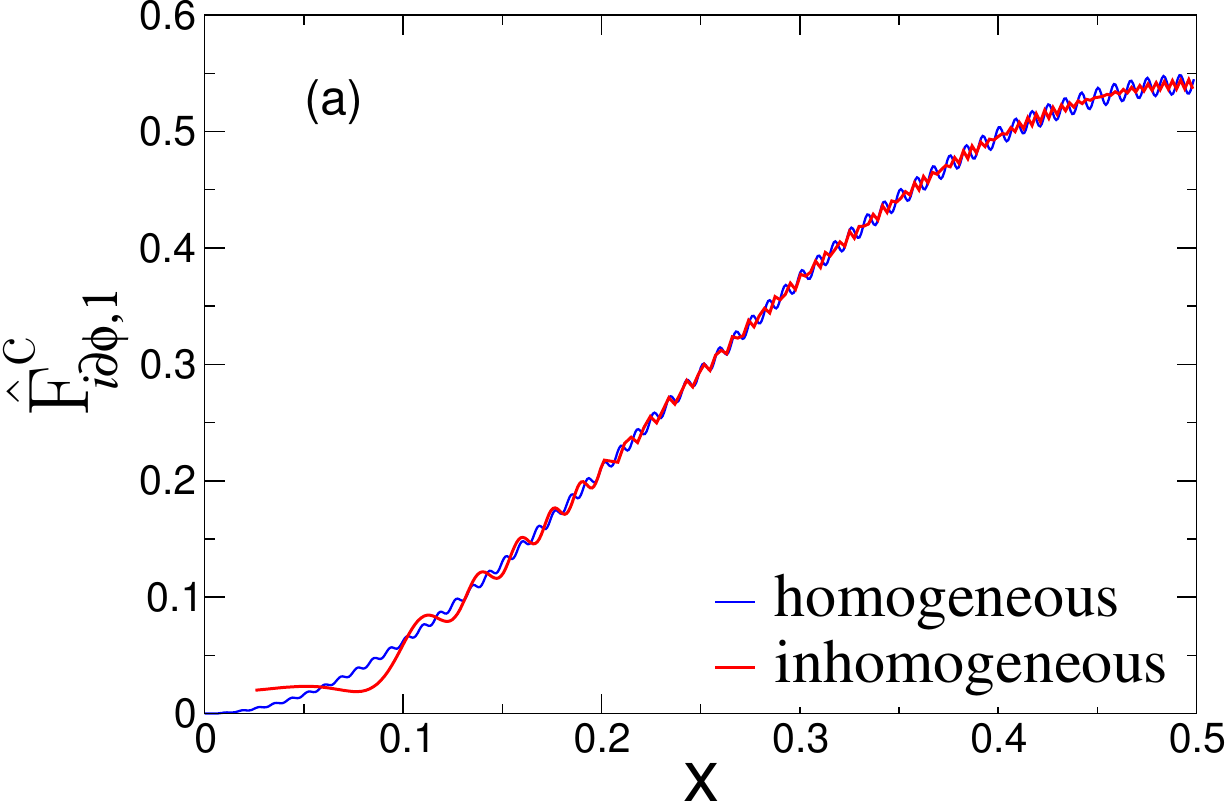}}
\subfigure
   {\includegraphics[width=0.45\textwidth]{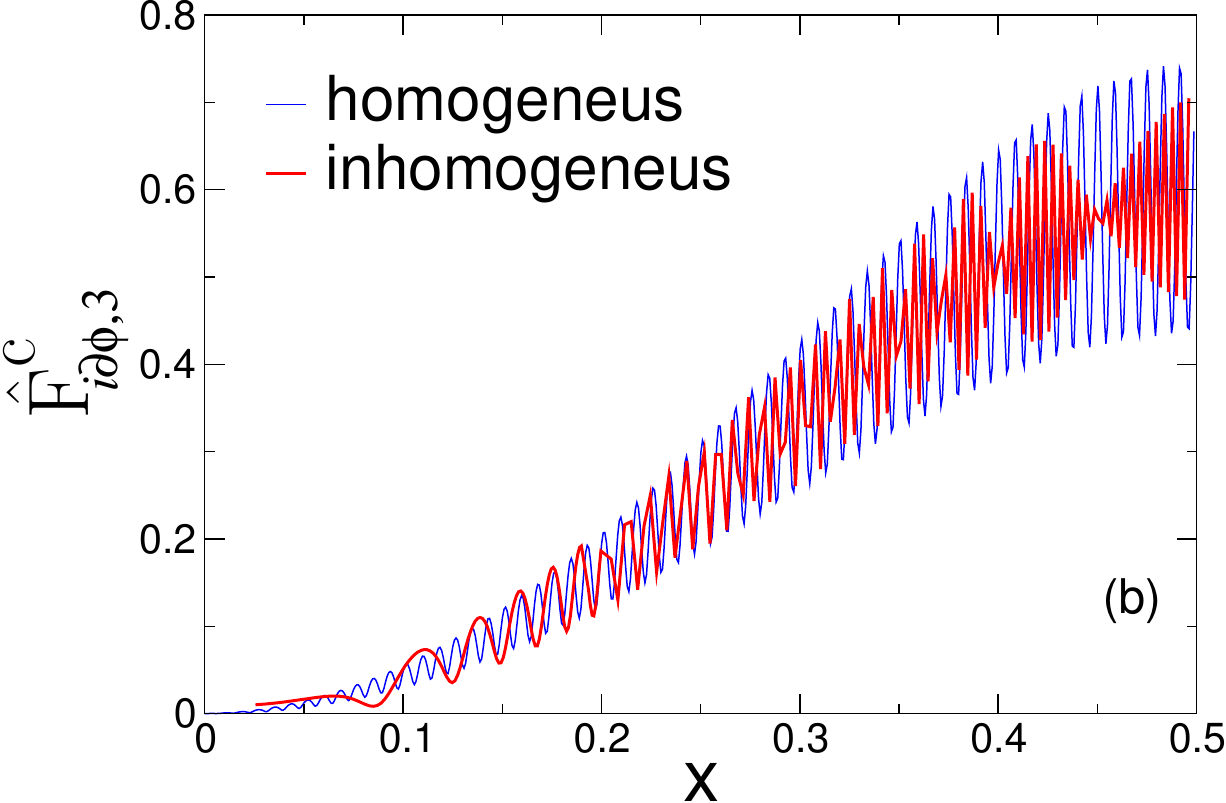}}
\caption{The same data as in Fig. \ref{fig:continuumm2}, but using as scaling variable ${\rm x}$ for the homogeneous gas and ${\rm x}'$ (cf. Eq. \eqref{eq:newcut}) for the 
inhomogeneous one. The smooth parts of the  two curves collapse on top of each other, as predicted by CFT. 
Conversely, the oscillations have very different features, in spite of the comparable amplitudes.
}
\label{fig:continuumm3}
\end{figure}

We finally discuss the corrections to the scaling which strongly affect the data in Figures \ref{fig:overlapVertex} and \ref{fig:continuumm1}.
Similar corrections have been found already for the ground state of the trapped gas \cite{random,vicari} and there is no quantitative understanding of them yet. 
The main reason for this lack of comprehension  is that there are at least two different sources of corrections which are intertwined in the final result
and disentangling them appears very complicated. 
First, also homogeneous systems in the ground state present oscillating deviations from the conformal asymptotic entanglement entropy which are called {\it unusual corrections}.
These have been characterised both in CFT \cite{sublead,ot-15}, and in microscopic models \cite{Fagotti2010cc,parity2,parity,correzioni,mint4,overlap1,num}.
They are unusual in the sense that the exponent governing their decay depends on the order $n$ of the R\'enyi entropy and it is not related to the 
leading irrelevant operator, like for standard corrections to the scaling. Indeed, it has been shown that these deviations decay as $\ell^{-\Delta/n}$ or $\ell^{-2\Delta/n}$,
for open and periodic systems respectively, where $\Delta$ is the scaling dimension of a {\it relevant} operator located at the conical singularities of the 
Riemann surface $\mathcal{R}_n$ \cite{sublead,ot-15}. 
Corrections with the same exponents have been found also for excited states of homogenous systems \cite{sublead1} (as expected since the structure of 
conical singularities does not depend on the state), but the amplitude depends in a very complicated and yet unknown manner on the state itself. 
These unusual corrections share many similarities with the ones observed in Figures \ref{fig:overlapVertex} and \ref{fig:continuumm1}, in particular they are larger for larger $n$
and very small at $n=1$. However, in the trapped case, the period of the oscillations depends on the position (not surprisingly since the density does) in a yet unclear fashion. 
A first naive guess for these corrections deep in the bulk might be that they are the same as in the homogenous systems, but with the density (i.e. $k_F(x)$) replaced by 
the local one. If this would be true the data for $\hat F_{i\partial \phi,n}({\rm x})$ and those for $\hat F^c_{i\partial \phi,n}({\rm x})$ should be very similar 
when the former is written as a function of ${\rm x}$ and the latter of ${\rm x}'$, because the mapping between the two makes the system homogenous.
In Fig. \ref{fig:continuumm3}, we explicitly perform this comparison, showing that deep in the bulk, while the amplitude of the oscillations is comparable, 
their periodicity  has a different structure.

A second correction  originates from the edges of the trapped system (at $\pm L$) and is strictly related to its inhomogeneities. 
Indeed, at finite $N$, near the edges, the fermion propagator deviates from thermodynamics form~\eqref{eq:gs2}:
in the proper subleading scaling variable, it can be expressed in terms of the Airy kernel, see e.g. \cite{random,e-13}. 
From the figures, in particular comparing with the results for the homogenous case in Figs. \ref{fig:continuumm2} and \ref{fig:continuumm3}, it is evident that close to the 
edge at ${\rm x}=0$, there are larger deviations than in the bulk and these are entirely due to the edge physics. 
These corrections right at the edge have been analytically worked out in \cite{e-13,random}, but it is not known how they get modified when entering the bulk.

\begin{figure}[t]
\centering
\subfigure
  {\includegraphics[width=0.4\textwidth]{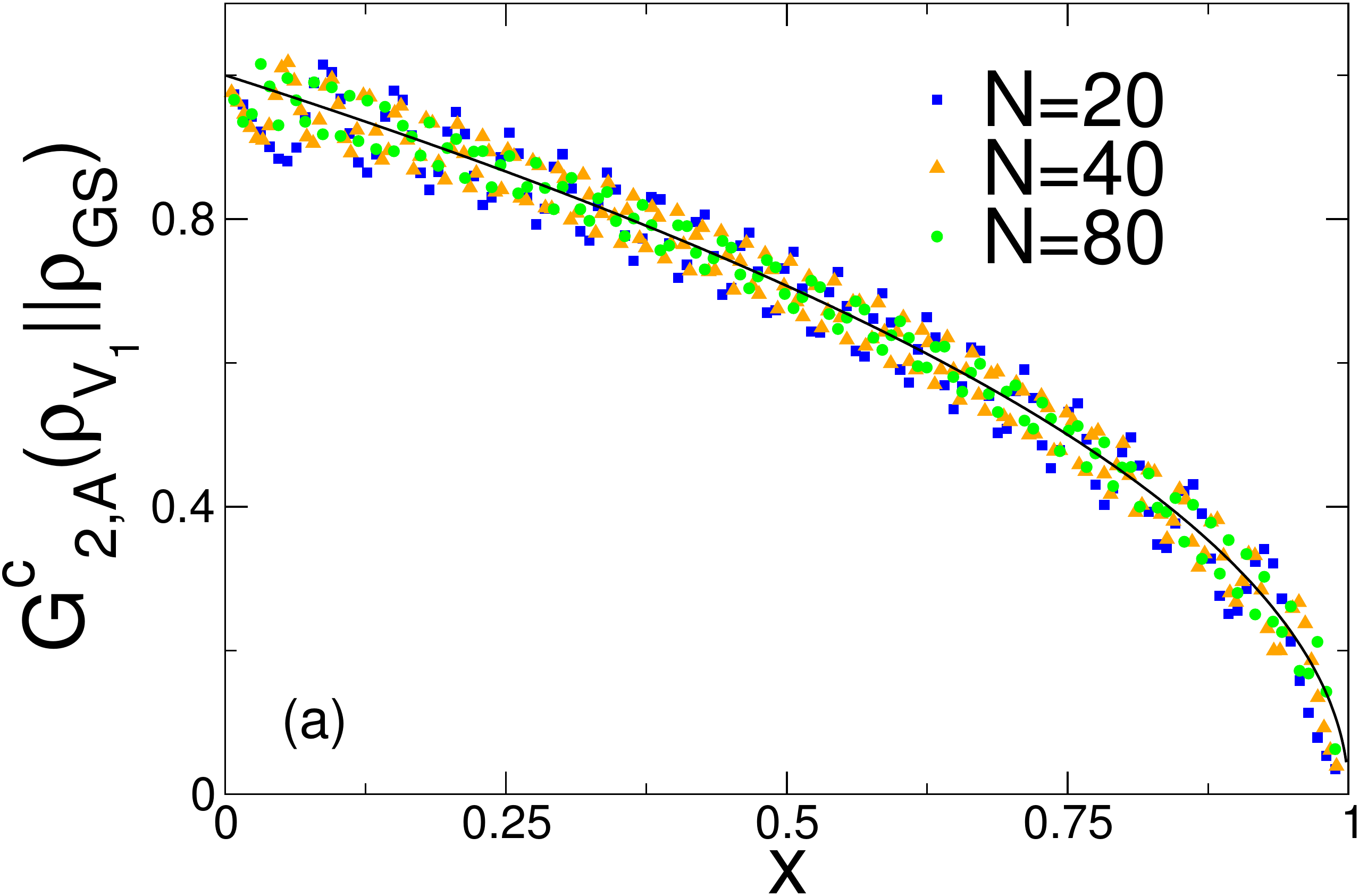}}
\subfigure
   {\includegraphics[width=0.4\textwidth]{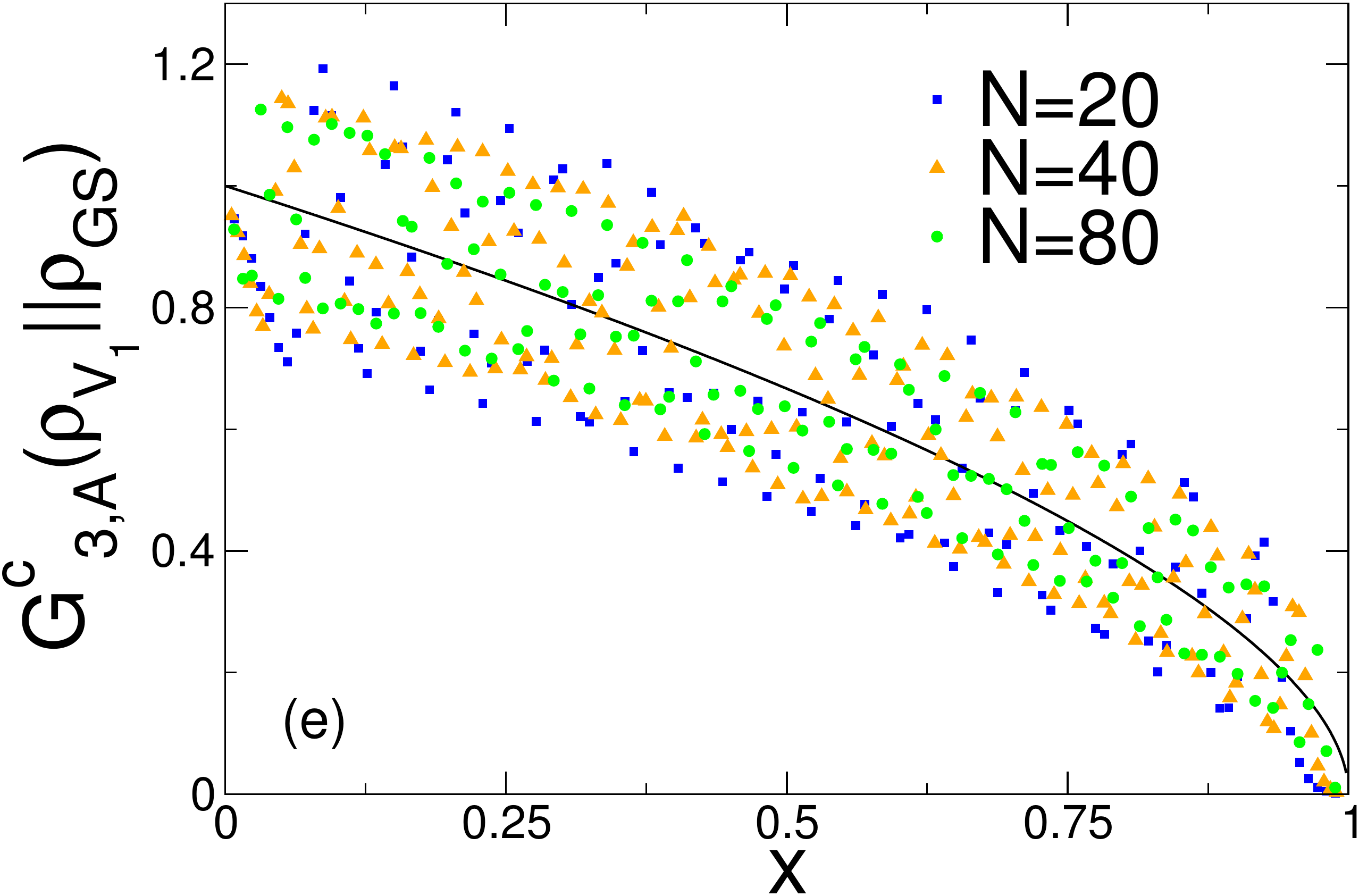}}
\subfigure
   {\includegraphics[width=0.4\textwidth]{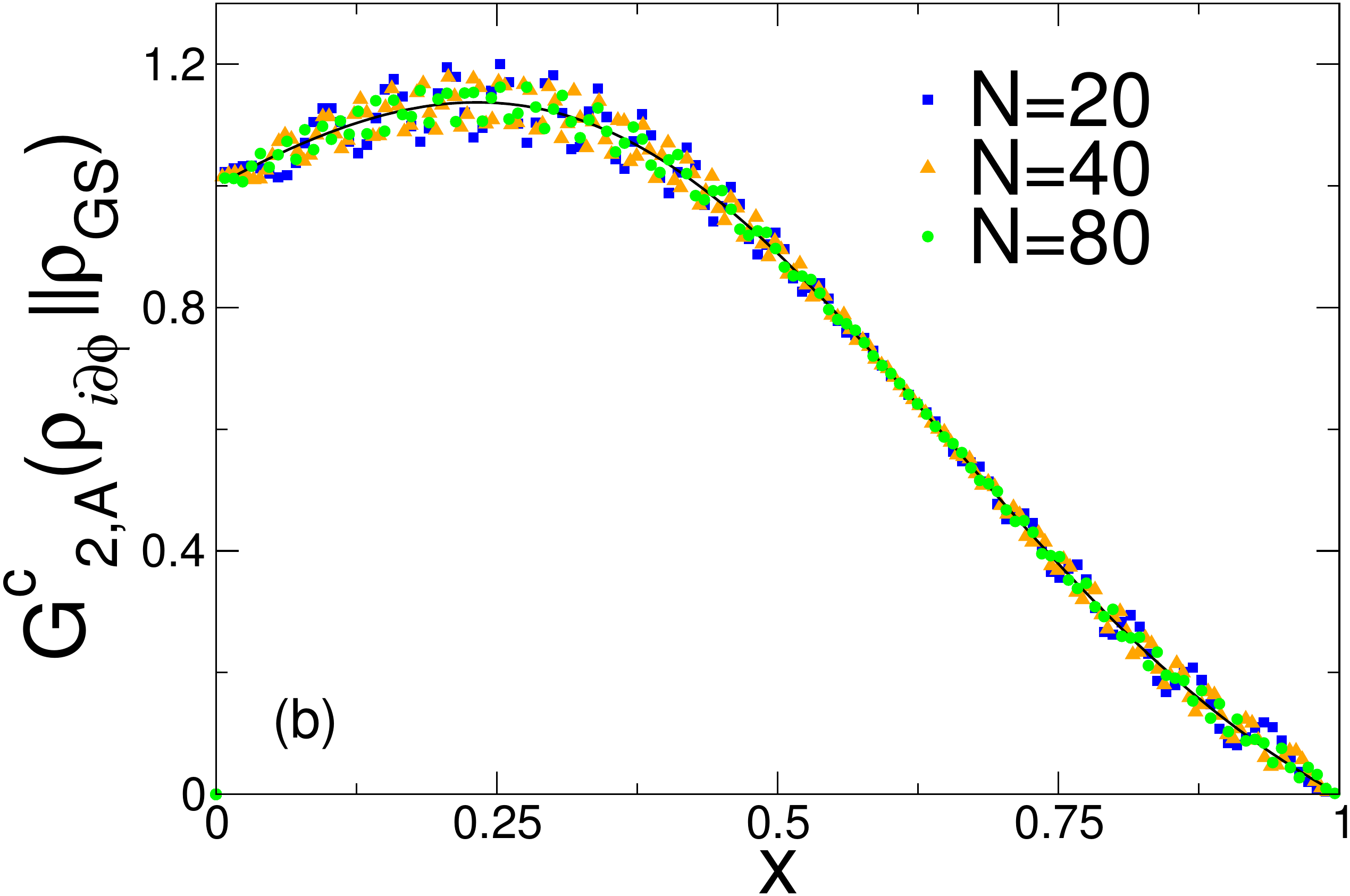}}
    \subfigure
     {\includegraphics[width=0.4\textwidth]{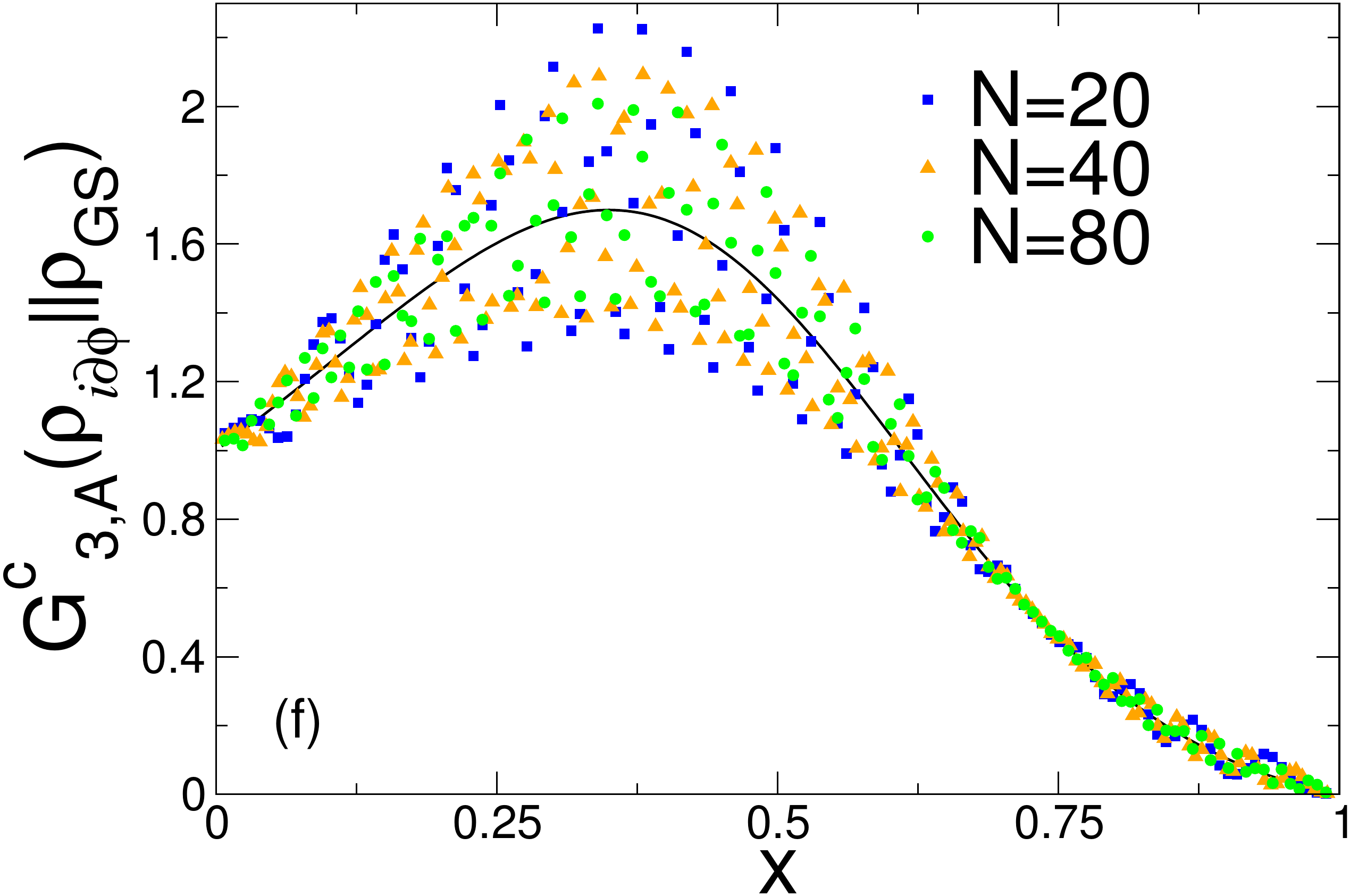}}
    \subfigure
     {\includegraphics[width=0.4\textwidth]{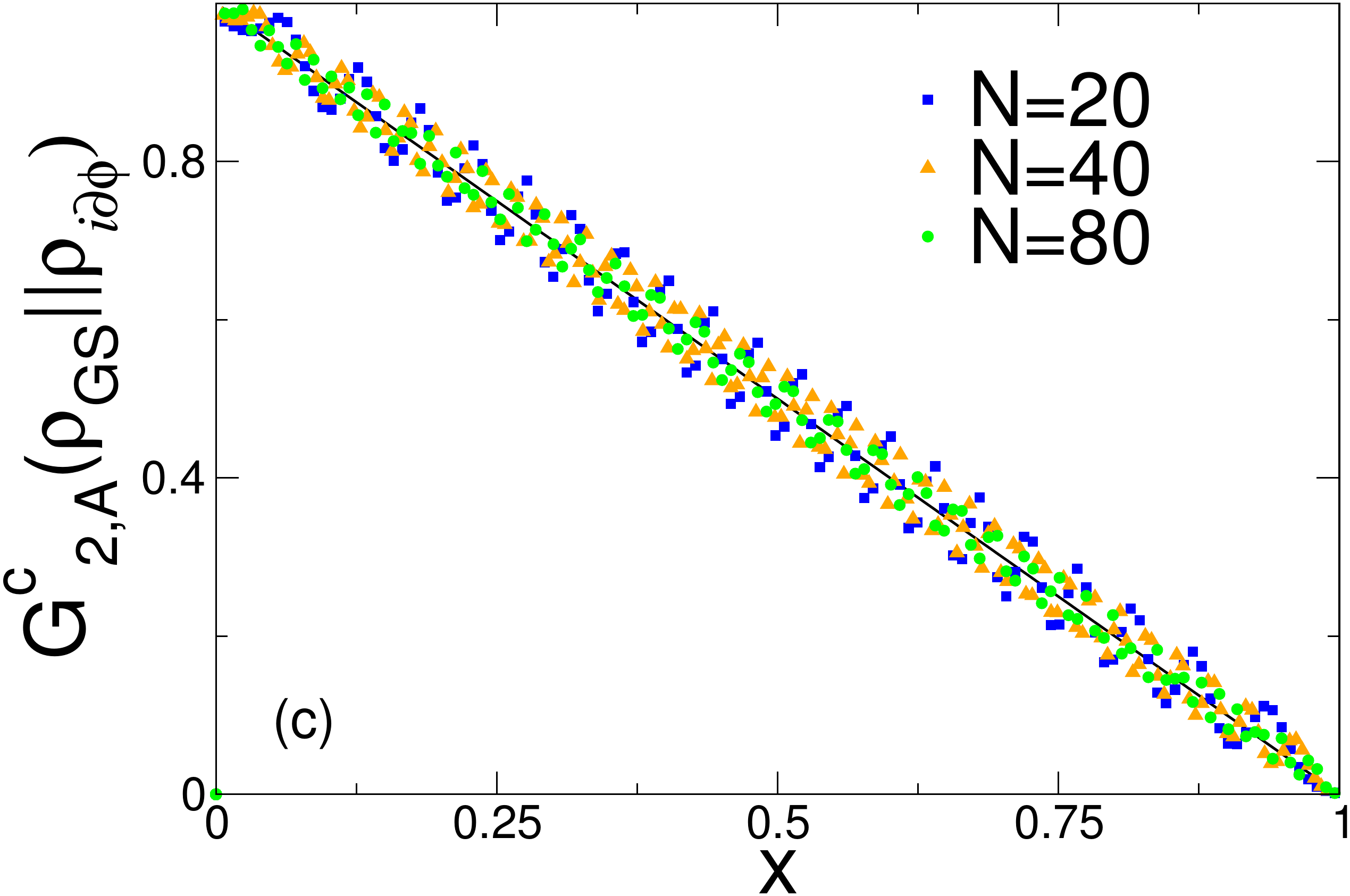}}
    \subfigure
      {\includegraphics[width=0.4\textwidth]{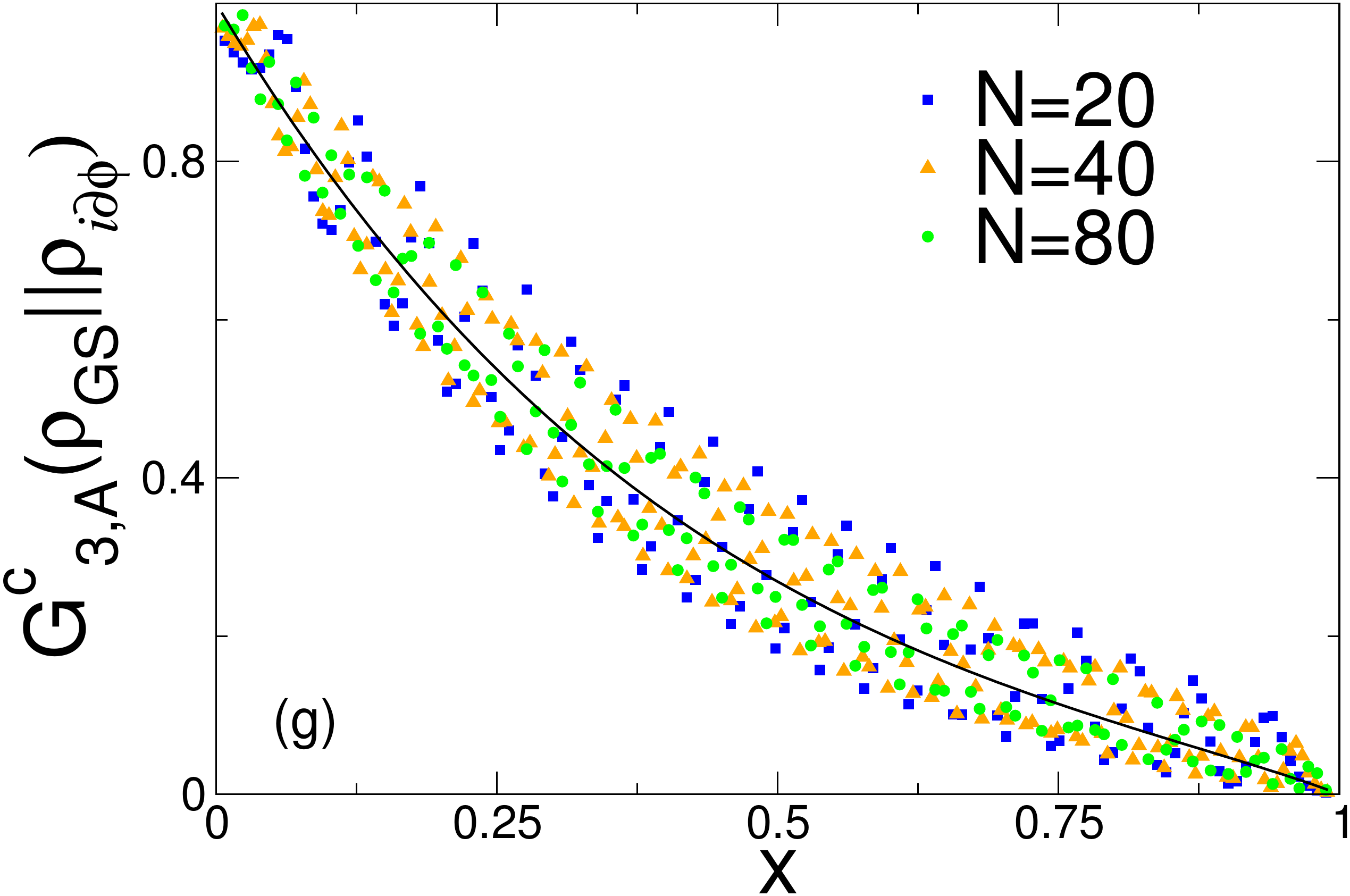}}
    \subfigure
  {\includegraphics[width=0.4\textwidth]{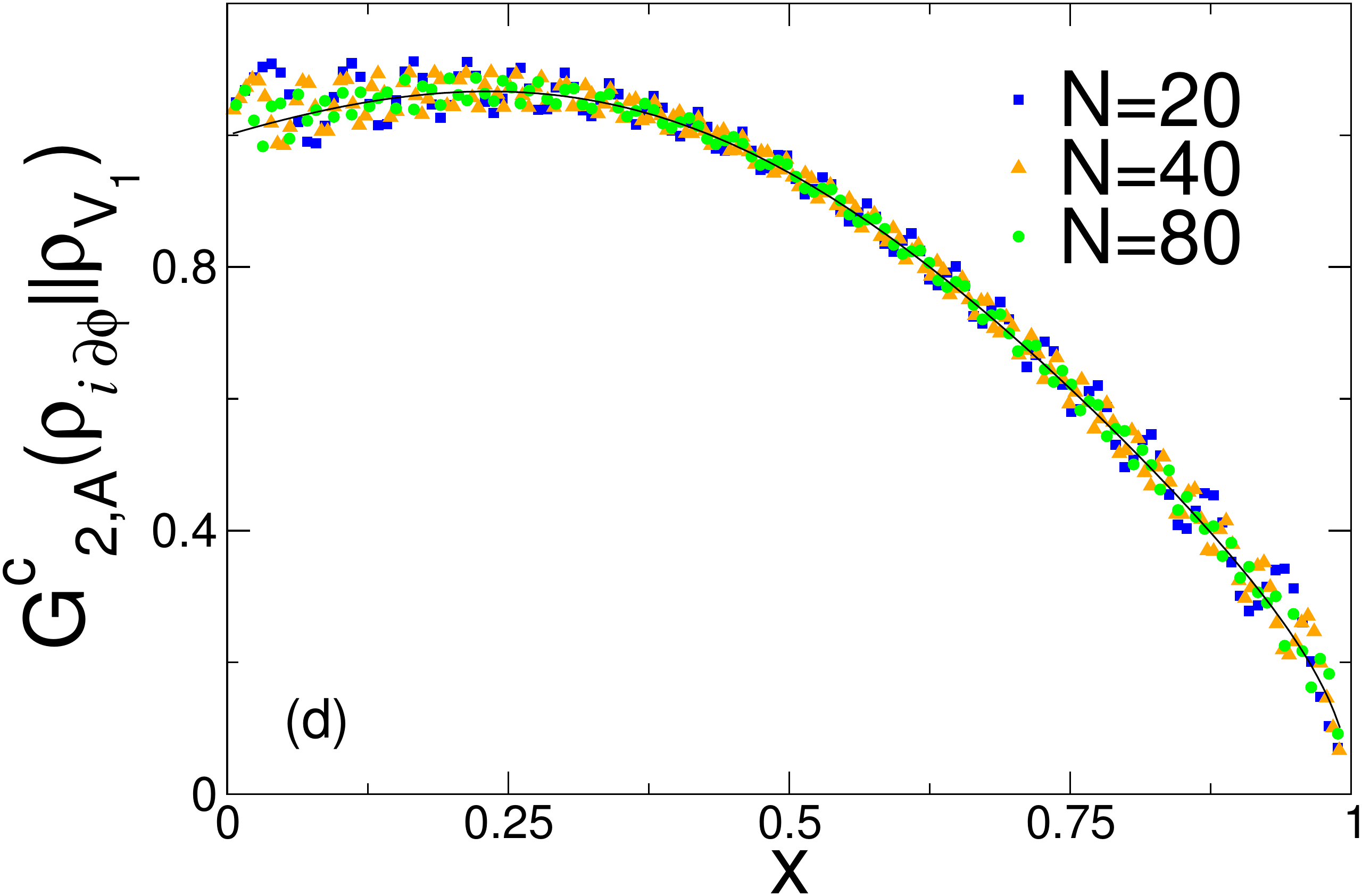}}
    \subfigure
  {\includegraphics[width=0.4\textwidth]{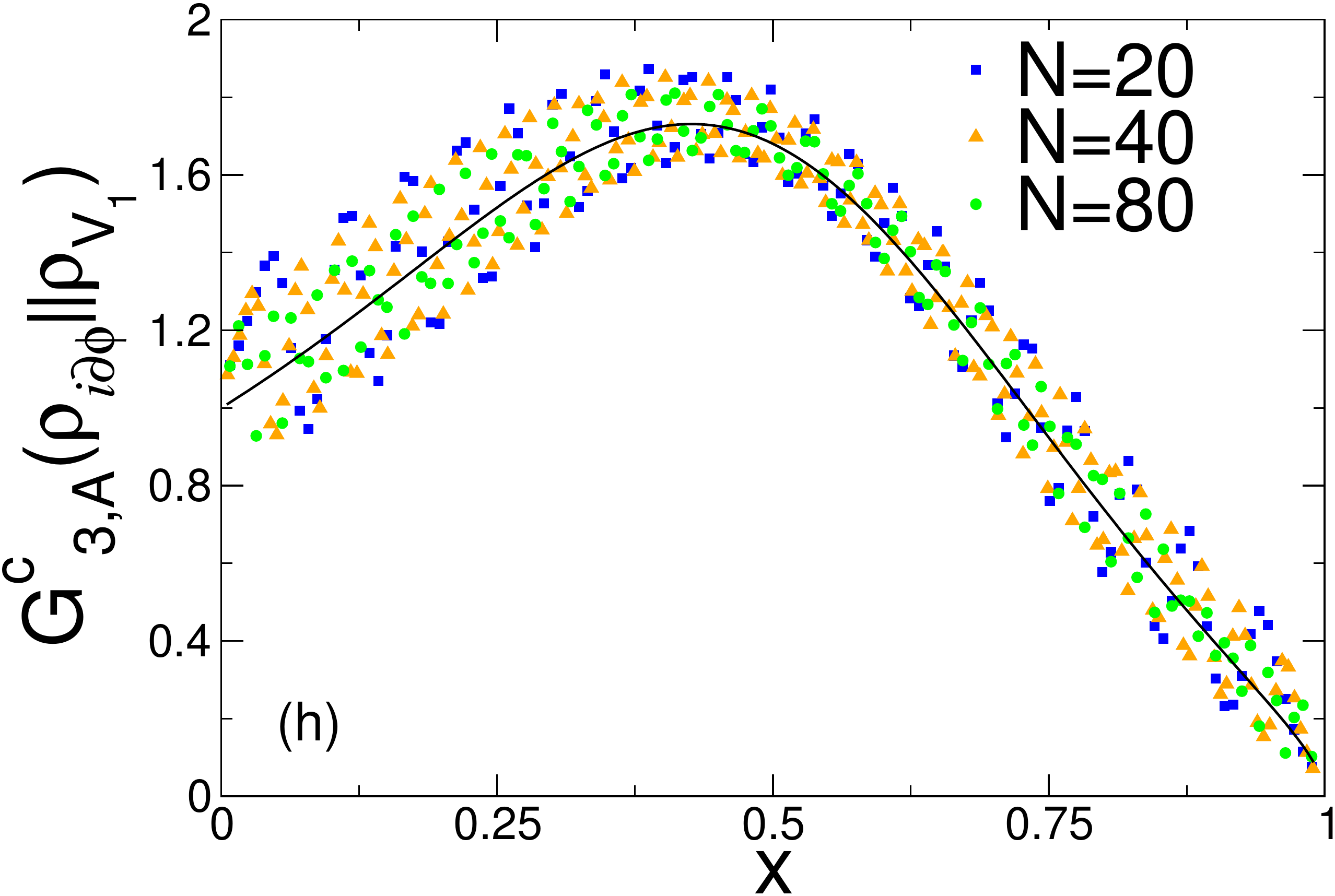}}
\caption{The ratio $G^c_{n, A} (\rho_1 \| \rho_0)$, Eq. \eqref{eq:univrelc}, as a function of ${\rm x}$ for $n=2,3$, for different pairs of states.
The considered states may be read off from the label of the vertical axis. 
The symbols are the exact numerical data at finite number of particles $N$. Different colours correspond to different $N$. 
The continuous curves are the CFT predictions in Eq. \eqref{eq:nuoveRel}. 
The data have been obtained with the Gauss--Legendre discretisation method, \ref{appendixD}.
}
\label{fig:nuoveRel}
\end{figure}

\subsection{Relative entropies}

In this subsection, we come to the R\'enyi relative entropies. Exploiting the techniques of \ref{appendixD} and \ref{appendixGamma}, we numerically compute the ratio
\begin{equation}
G^c_{n, A}(\rho_1||\rho_0)=\dfrac{\mathrm{Tr}(\rho_1\rho_0^{n-1})}{\mathrm{Tr}\rho_1^n},
\end{equation}
for $n=2, 3$ and with $\rho_0$ and $\rho_1$ corresponding to the possible pairs of  states considered in the previous subsection, i.e. ground-state, particle-hole excitation,
addition of a particle. 
In the thermodynamic limit the numerical data are expected to converge to the CFT predictions for $G^c_{n, A}(\rho_1||\rho_0)$ in Eq. \eqref{eq:nuoveRel}. 
The numerical results are shown in Figure \ref{fig:nuoveRel} for different number of particles $N$ and for all possible pairs of states.

It is evident from all data in Fig. \ref{fig:nuoveRel} that increasing $N$ the numerics converge toward the asymptotic CFT predictions. 
Actually, the overall agreement is exceptionally good given the presence of oscillating corrections to the scaling (which, incidentally, appear to be  larger than 
those in the homogeneous case, see \cite{Paola1}). 
Exactly like for the R\'enyi entropies in the previous subsection, these corrections have at least a twofold origin:  
i) the geometrical structure of the Riemann surface defining $G_{n, A}(\rho_1||\rho_0)$ is the same as the one for the R\'enyi entropies, in particular with the same 
conical singularities; hence the same kind of unusual corrections are expected;
(ii) close to the edge, there are subleading corrections to the two-point function which affect both the density matrices $\rho_0$ and $\rho_1$ and hence the relative entropies. 
Unfortunately, as for the R\'enyi entropies, it is not possible to disentangle these effects and have a quantitative descriptions of these oscillating deviations 
from the asymptotic behaviour, as instead done for the homogeneous case \cite{Paola1}.

\section{Discussion and outlook} \label{sec:conclusions}

This work is set within the context of the study of entanglement in inhomogeneous many-body quantum systems and its relation with conformal field theory.
The core of this new line of research lies in the fact that the long distance behaviour of such systems may, under given assumptions, be described by CFT in a curved spacetime whose metric encodes the inhomogeneity parameters.
In particular, here we provided new analytical predictions for the R\'enyi and relative entanglement entropies of low-energy excitations in the inhomogeneous one-dimensional free Fermi gas, for an interval adjacent to the physical edge (i.e. where the density of particles vanishes, $\rho (x)=0$). 
Our main analytical results are given by equations \eqref{eq:nuove}, \eqref{analyti} for the universal ratio $F^c_{\Upsilon, n} (A)$ and by equations \eqref{eq:nuoveRel}, 
\eqref{analityRel} for $G^c_{n} (\rho_1 \| \rho_0)$. These predictions have been tested against exact numerical data, finding perfect agreement.




Starting from this work, several directions would be worth investigating.
First,  we only considered excitations associated to primary fields. It would be interesting to generalise  our findings to a generic excited state. For homogeneous systems this generalisation  was considered in Refs. \cite{palmai, descendants1}. However, as soon as descendent fields are involved, the calculations become much more cumbersome.
It is also quite natural to ask whether our formalism could be generalised to interacting problems, e.g., to some inhomogeneous version of the XXZ spin chain
or Lieb-Liniger gas, as e.g.  in \cite{eisler}. 
The main problem is that for interacting systems the inhomogeneity affects two parameters:
in addition to the metric,  also the Luttinger parameter $K$, related to the couplings of the microscopic Hamiltonian \cite{giamarchi}, varies in space. 
Fixing these parameters is not an easy task, but, more importantly, this space-dependent parameter $K$  turns out to break  conformal invariance \cite{brune}, 
so that we have to deal with a more complicated theory.

\section*{Acknowledgments}

The authors would like to thank Vincenzo Alba and Giuseppe Di Giulio for useful discussions.
PC acknowledges support from ERC under Consolidator grant  number 771536 (NEMO). 
Part of this work has been carried out during the workshop  ``Entanglement in quantum systems'' at the Galileo Galilei Institute (GGI) in Florence.

\begin{appendix}

\section{Numerical tools} \label{appendixTools}

In this appendix we describe how the numerical data reported in Section \ref{sec:nchecks} have been obtained. 

Consider a gas of $N$ non-interacting spinless fermions trapped by an external potential.
The two-point correlation function is 
\begin{equation}
\label{eq:one-p}
C(x,y) \equiv \braket{c^{\dagger}(x)c(y)}=\sum_{k=1}^N\phi_k^*(x) \phi_k(y),
\end{equation}
where $c(x)$ ($c^\dag(x)$) is the fermionic annihilation (creation) operator and $\{ \phi_k (x) \}$ represents a complete set of the (normalised) eigenfunctions 
of the one-particle problem. 
Using the Christoffel-Darboux formula  the correlation function may be rewritten as 
\begin{equation}
\label{eq:kernelGUE}
C(x,y)=
\begin{cases}
\sqrt{\dfrac{N}{2}} \dfrac{\phi_{N+1}(x)\phi_N(y)-\phi_N(x)\phi_{N+1}(y)}{x-y}, \qquad  &x\neq y, \\
\phi'_{N+1}(x)\phi_N(x)-\phi'_N(x)\phi_{N+1}(x),\quad &x= y.
\end{cases}
\end{equation}
In the special case of the harmonic confining potential, $C(x,y)$ may be rewritten in terms of the Hermite polynomials, but the precise 
expressions are unimportant. 

The reduced density matrix restricted to $A=[x_1,x_2]$ has the Gaussian form
\begin{equation}
\label{eq:redA}
\rho_A \propto \exp -\int_{x_1}^{x_2} dy_1dy_2 c^{\dagger}(y_1)\mathcal{H}(y_1,y_2) c(y_2) ,
\end{equation}
where the entanglement Hamitonian ${\cal H}$ in terms of the reduced correlation matrix $C_A(x,y)=P_A C P_A$ ($P_A$ is the projector on the interval $A$) is 
\begin{equation}
\mathcal{H}=\log [(1-C_A)/C_A].
\label{HvsC}
\end{equation}
This last relation is the continuum limit of the analogous one in the fermionic lattice model (see Refs. \cite{peschel2001, peschel2003}). 
Eqs. \eqref{eq:redA} and \eqref{HvsC} imply that the (R\'enyi) entanglement entropies may be written in terms of the $N$ non-zero eigenvalues $\lambda_i$ of $C_A(x,y)$ as
\begin{equation}
S^{(n)}_{A}= \sum_{i=1}^N e_n(\lambda_i)\quad {\rm with}\quad e_n(\lambda)=\dfrac{1}{1-n} \log [\lambda^n+(1-\lambda)^n].
\label{eq:def}
\end{equation}
In the following we review some well known techniques to  extract  effectively the eigenvalues of the reduced correlation matrix   in continuous models.

\subsection{Overlap matrix method.} \label{appendixOM}

The first method we  describe allows us to obtain the eigenvalues of the reduced correlation matrix (and hence 
R\'enyi entropies) from an $N\times N$ matrix, known as overlap matrix, which shares its non-zero eigenvalues with $C_A(x,y)$ \cite{vicari,overlap1,overlappp}. 
This technique has already been used for several applications in the context of the 
entanglement \cite{vicari,mint1,mint2,mint3,mint4,si-13,ep-13,nv-13,csc-13,v-12,ckc-14,o-14,hm-15,cw-16,b-17,lms-18}. %
 
The starting point is that the characteristic  polynomial of $C_A(x,y)$ is a \emph{Fredholm determinant} 
\begin{equation}
\label{eq:fredh}
D_A(\lambda)=\det [\lambda \delta_A(x,y)-C_A(x,y)],
\end{equation}
where $ \delta_A(x,y)=P_A\delta(x-y)P_A$. 
The  operative definition of this determinant is 
\begin{equation}
\label{eq:contin1}
\log \, \det [\lambda \delta_A(x,y)-C_A(x,y)]=-\sum_{k=1} ^{\infty} \dfrac{\lambda^{-k} \mathrm{Tr}\,C_A^k}{k},
\end{equation}
where the traces are
\begin{equation}
\label{eq:traces}
\mathrm{Tr}\,C_A^k=\int \, dx_1dx_2 \dots dx_k C_A(x_1,x_2)C_A(x_2,x_3) \dots C_A(x_{k-1},x_k)C_A(x_k,x_1).
\end{equation}

Since the  zeroes of $D_A(\lambda)$ are the eigenvalues of $C_A$, making use of the Cauchy residue theorem, the R\'enyi entanglement entropies 
may be written as the contour integral 
\begin{equation}
\label{eq:integral}
S^{(n)}_A=\oint_{\mathcal{C}} \dfrac{d\lambda}{2\pi i} e_n(\lambda)\dfrac{d \log D_A(\lambda)}{d \lambda},
\end{equation}
where $e_n(\lambda)$ is defined in \eqref{eq:def} and $\mathcal{C}$ is a contour which encircles the segment $(0,1)$--where the eigenvalues of $C_A$ lie.

The computation of the Fredholm determinant is simplified by the introduction of the $N \times N$ \emph{overlap matrix} $\mathbf{A}$ with elements
\begin{equation}
\label{eq:overlap}
\mathbf{A}_{nm}\equiv\int_{x_1}^{x_2} dz \, \phi^*_n(z) \phi_m(z).
\end{equation}
It is immediate to show that 
\begin{equation}
\label{eq:overlapC}
\begin{split}
\textrm{Tr}  \mathbf{A}^k =\textrm{Tr} C_A^k &,
\end{split}
\end{equation}
and hence the overlap matrix $\mathbf{A}$ and the reduced correlation matrix $C_A(x,y)$ have the same non-zero eigenvalues. 
Thus, the final  result is 
\begin{equation}
\label{eq:integral1}
S^{(n)}_A= \sum_{m=1}^N e_n(a_m).
\end{equation}
where $a_m$ are the eigenvalues of $\mathbf{A}$.

\subsection{Gauss--Legendre discretisation  of continuous matrices.} \label{appendixD}

The eigenvalues of  the reduced correlation matrix may be obtained directly by exploiting effective discretisations of the kernel  
as those proposed in \cite{bornemann, bornemann1}.  

In the continuous case, the eigenvalue problem for the reduced correlation matrix $C_A (x, y)$ is defined by the following integral equation
\begin{equation} \label{eq:eigC_A}
\int_{x_1}^{x_2} dy \, C_A (x, y) \psi (y) = \lambda \psi (x).
\end{equation}
Let us introduce the Nystr\"om'€™s idea of classical quadrature to solve \eqref{eq:eigC_A}. The quadrature rule
\begin{equation}
\label{eq:quadratura}
P(f)=\sum_{j=1}^mw_jf(x_j) \approx \int_{x_1}^{x_2}f(x) \, dx,
\end{equation}
states that the integrand is evaluated at a finite set of $m$ points called integration points and a weighted sum of these values is used to approximate the integral. We are going to use the Gauss-Legendre rule, which is a family of quadrature rules with positive weights. 
We first trivially convert the limits of integration for the interval $[x_1,x_2]$ to the Gauss-Legendre interval $[-1,1]$
\begin{equation}
\begin{split}
\label{eq:gauss-leg}
\int_{x_1}^{x_2}  dx\,f(x) &=\dfrac{x_2-x_1}{2} \int_{-1}^1  dx \,f \left( \dfrac{x_2-x_1}{2}x+\dfrac{x_2+x_1}{2} \right) \, \\
& \approx \dfrac{x_2-x_1}{2}\sum_{j=1}^mw_if \left( \dfrac{x_2-x_1}{2}x_j+\dfrac{x_2+x_1}{2} \right). 
\end{split}
\end{equation}
In the Gauss-Legendre quadrature rule, the points $x_j$ in Eq. \eqref{eq:gauss-leg} are chosen to be the roots of the Legendre polynomial. 
Given this rule, the discretised version of \eqref{eq:eigC_A} is 
\begin{equation}
\label{eq:fredholm1}
\sum_{j=1}^m w_j C(x_i,x_j) \psi_j= \lambda \psi_i,
\end{equation}
which has to be solved for $\psi_i \approx \psi(x_i),$  $  i=1,\dots,m$. 
Thus we only need to diagonalise the $m\times m$ matrix $w_j C(x_i,x_j)$, or, in a more symmetric, but equivalent, fashion
\begin{equation}
\label{eq:Cij}
C_{ij} \equiv w_i^{1/2} C(x_i,x_j) w_j^{1/2} .
\end{equation}
The numerical implementation of this method is straightforward.

\subsection{Product of Gaussian operators} \label{appendixGamma}

When computing the R\'enyi relative entropies between the reduced density matrices of two different eigenstates, we need to evaluate objects of the following form
\begin{equation}
\label{eq:goal}
\mathrm{Tr}(\rho_1 \rho_0^{n-1}).
\end{equation}
Generically speaking, the two RDMs $\rho_1$ and $\rho_0$ do not commute and so they cannot be simultaneously diagonalised to calculate the relative entropies from their eigenvalues in a common basis \cite{Paola1}. 
However, it is possible to use the composition properties of gaussian density matrices to compute the traces of the product of a generic number of them. 

Let $\rho [\Gamma]$ denote a gaussian density matrix characterised  by its correlation function $\Gamma$, written in terms of Majorana fermions. $\Gamma$ is related to the correlation matrix $C$ as
\begin{equation}\label{eq:fact}
\Gamma=(\mathbb{I}-2 C) \oplus (2 C^{\mathrm{T}}-\mathbb{I}).
\end{equation}
In the continuous model, since $C(x, y)$ is continuous, so is $\Gamma (x, y)$. However, as just shown, $C$ can be discretised, Eq. \eqref{eq:Cij}, so that from \eqref{eq:fact} we get a discretised  version of $\Gamma$, $\Gamma_{ij}$.
In this way, we trace the problem back to a discrete one, which has been considered in Ref. \cite{fagottiXY}, where the algebra of gaussian RDMs has been worked out. Below we briefly review the main points.

The composition of correlation matrices is indicated by $\Gamma \times \Gamma'$ (not to be confused with the product of the matrices) and it is implicitly defined by
\begin{equation}
\label{eq:product}
\rho[\Gamma]\rho[\Gamma']= \mathrm{Tr}[\rho[\Gamma]\rho[\Gamma']]\rho[\Gamma \times \Gamma'],
\end{equation}
leading to
\begin{equation}
\label{eq:productrule}
\Gamma \times \Gamma'=1-(1-\Gamma')\dfrac{1}{1+\Gamma \Gamma'}(1-\Gamma).
\end{equation}
The trace of two fermionic operators can be computed as
 \begin{equation}
 \label{eq:productrule1}
 \{ \Gamma, \Gamma' \} \equiv \mathrm{Tr}(\rho_{\Gamma}\rho_{\Gamma'})=\prod_{\mu \in \mathrm{Spectrum} [\Gamma \Gamma']/2}\dfrac{1+\mu}{2},
 \end{equation}
namely  it is the product of the eigenvalues of ($1 + \Gamma \Gamma')/2$ with halved degeneracy.
Once we have this explicit result for the product of two gaussian operators, the trace of the product of a generic number of them can be computed iteratively as follows
\begin{eqnarray} \label{iteration}
 \{ \Gamma_1, \cdots, \Gamma_n \} \equiv {\rm Tr} \left( \rho_{\Gamma_1} \cdots \rho_{\Gamma_n} \right)  = \{ \Gamma_1 , \Gamma_2 \} \{ \Gamma_1 \times \Gamma_2, \Gamma_3, \cdots , \Gamma_n \}.
\end{eqnarray}

\end{appendix}

\end{document}